\documentclass[letterpaper]{article}

\usepackage{enumitem}
\usepackage[T1]{fontenc}
\usepackage{xcolor}
\usepackage[utf8]{inputenc}
\usepackage{geometry}
\usepackage{newunicodechar}
\newunicodechar{′}{\ensuremath{'} }
\newunicodechar{̀}{\`{}}
\usepackage{setspace}
\usepackage{lineno}

\usepackage[
    backend=biber,
    style=chem-acs,
    subentry,
    articletitle=true
]{biblatex}
\usepackage{amsmath}
\usepackage{graphicx}
\usepackage{float}
\usepackage{booktabs}

\usepackage{siunitx}
\DeclareSIUnit\Molar{M}
\DeclareSIUnit\rpm{rpm}
\usepackage[version=4]{mhchem}

\newfloat{scheme}{htbp}{los}
\floatname{scheme}{Scheme}
\floatname{chart}{Chart}
\newfloat{graph}{htbp}{loh}

\usepackage{authblk}
\author[1]{Akeshi Aththanayake}
\affil[1]{Department of Physics, Case Western Reserve University, 10900 Euclid Ave, Cleveland, Ohio 44106, United States}
\author[1]{Andrew Lininger}
\author[1]{Khoi-Anh Pham}
\author[2,3]{Radu Malureanu}
\affil[2]{Department of Electrical and Photonics Engineering, Technical University of Denmark, 2800 Kongens Lyngby, Denmark}
\affil[3]{National Centre for Nano Fabrication and Characterization, Technical University of Denmark, 2800 Kongens Lyngby, Denmark}
\author[4]{Divita Mathur*}
\affil[4]{Department of Chemistry, Case Western Reserve University, 10900 Euclid Ave, Cleveland, Ohio 44106, United States}
\author[1,5,6]{Giuseppe Strangi*}
\affil[5]{Case Comprehensive Cancer Center, Case Western Reserve University, 10900 Euclid Ave, Cleveland, Ohio 44106, United States}
\affil[6]{NLHT Labs - Department of Physics and CNR-NANOTEC Istituto diNanotecnologia, University of Calabria, Via Pietro Bucci 87036, Rende, Italy.}
\title{Enhancing FRET through DNA-controlled Emitters and ENZ Metamaterials}
\date{*Email: divita.mathur@case.edu, giuseppe.strangi@case.edu}

\begin{document}
\begin{refsection}
\maketitle

\begin{abstract}
Simultaneous nanoscale spatial control and macroscopic electromagnetic enhancement are critical for maximizing resonance energy transfer. However, conventional systems with photonic platforms suffer from stochastic emitter distributions and steep field gradients. Here we demonstrate a unified architecture combining programmable DNA molecular beacons with an epsilon near zero (ENZ) multilayer metamaterial. Using DNA scaffolds, donor-acceptor pairs (Atto425/Cy3.5) are structurally locked into closed (2 nm) and extended (8.16 nm) conformations within the ENZ near field. Time-resolved photoluminescence reveals accelerated donor decay and an increase in extended state FRET efficiency from 56.7\% on glass to 79.4\% on ENZ, corresponding to a 5.8 fold transfer rate acceleration. Electromagnetic Green tensor modeling confirms that ENZ boundary conditions strongly amplify dipole coupling within first few nm of the interface exhibiting pronounced orientation selectivity. This hybrid platform provides a deterministic orientational and spatial control for sculpting molecular-scale light-matter interactions.
\end{abstract}


\noindent Manipulating light-matter interactions at the nanoscale is central to advancing quantum photonic and optical technologies. ~\supercite{novotny2012principles,lodahl2015interfacing,donato2026observation,aththanayake2025tunable,lininger2023chirality} 
While \emph{classical} systems tolerate microscopic disorder through ensemble averaging  ~\supercite{shannon1948mathematical} \emph{quantum} technologies rely on precisely controlled  resonance energy transfer, coherent dipole-dipole coupling, and collective emission between discrete emitters.~\supercite{kimble2008quantum,el2013resonant,chattaraj2025energy} These fundamental processes depend critically on the local electromagnetic environment and intermolecular spatial separation.\supercite{purcell1995spontaneous} Consequently, deterministic spatial positioning remains the primary lever for regulating these nanoscale interactions, posing a central challenge for scalable quantum architectures.\supercite{scully1997quantum,li2019resonance} 

Engineered nanophotonic substrates such as optical micro cavities, photonic crystals, and plasmonic nanostructures tailor these interactions by modifying the local density of optical states (LDOS).\supercite{ebbesen1998extraordinary,kuhn2006enhancement,bellessa2004strong,novotny2012principles,chikkaraddy2016single,de20133d,staude2013tailoring}
Macroscopic quantum electrodynamics (QED) and Green tensor formalisms have long established how dielectric and metallic interfaces reshape dipole decay rates and intermolecular energy transfer pathways.\supercite{gersten1981spectroscopic,sipe1987new, Dung2002} Subsequent studies confirmed that localized and surface plasmon resonances can dramatically alter energy transfer rates and enhance acceptor photoluminescence.\supercite{lakowicz2005radiative, andrew2004energy, hsu2017plasmon,schuller2010plasmonics,lunz2011surface} 
However, conventional dielectric and plasmonic environments typically produce extreme field gradients and localized hotspots.\supercite{koenderink2015nanophotonics} 
Similarly, hyperbolic metamaterials (HMMs), anisotropic media with opposite sign principal permittivity components, can modify FRET through high-\textit{k} modes. Experimental studies have demonstrated FRET enhancement \textit{inside} gold-nanorod type-I HMMs (up to 13x), over long distances through multilayer media ($\sim 160$~nm), and for ensembles above layered HMMs (1x-4x efficiency enhancements reported for different emitter systems)\supercite{Roth2018FRET,Deshmukh2018LongRange,Newman2018Dipole,Shih2023FRET,Chao2024FRET}.
In ensemble measurements, spatial variability couples with stochastic emitter distributions and dynamic dipole orientation,\emph{smearing} molecular-scale interactions and yielding suboptimal, unreproducible performance.\supercite{hoang2016ultrafast,curto2010unidirectional}

Epsilon-near-zero (ENZ) metamaterials offer a compelling route to overcome these spatial limitations.~\supercite{kiasat2023epsilon,sreekanth2014large} When the real part of the effective permittivity approaches zero, the effective wavelength diverges and phase advance is suppressed, generating spatially uniform electromagnetic fields across extended planar regions.~\supercite{liberal2017near,reshef2019nonlinear} Unlike previous platforms including resonant cavities, localized plasmonic hotspots, and HMMs\supercite{konrad2015controlling,buendia2024long,lal2007nano,bouchet2016long,ghenuche2014nanophotonic} thin-film ENZ platforms provide broad-area LDOS enhancement in a slow light regime with minimal dephasing.~\supercite{xie2025resonant,stengel2026quantum,de2014double,infusino2014loss,krishna2016dye,jaffray2022near} 
Crucially, ENZ media are theoretically predicted to enhance and extend resonance transfer beyond the conventional $1/r^6$ distance scaling.~\supercite{deshmukh2018long,wu1994resonance}

Nevertheless, capitalizing on macroscopic ENZ enhancement requires precise regulation of emitter geometry at the molecular scale. ~\supercite{koenderink2015nanophotonics} DNA nanotechnology provides an ideal solution to this geometric challenge.~\supercite{mathur2020can,adamczyk2022dna,scott2025transport,mathur2023pursuing} Synthetic DNA tile assemblies and molecular rulers enable sub-nanometer emitter positioning, drastically reducing the stochasticity inherent in colloidal platforms. \supercite{cannon2015excitonic,cervantes2022single,hemmig2016programming,acuna2012fluorescence,rothemund2006folding,liu2011three,tyagi1996molecular} DNA-templated assembly has also enabled the development of sophisticated hybrid nanophotonic systems, leveraging precise spatial control to engineer plasmonic resonances and enhance light-matter interactions at the nanoscale. ~\supercite{govorov2010theory, kuzyk2012dna} Furthermore, DNA scaffolds restrict the rotational diffusion of conjugated fluorophores, suppressing orientation factor ($\kappa^2$) variability and mitigating ensemble broadening.\supercite{kroutil2014cy3,stennett2014photophysical,iqbal2008orientation} In particular, simple DNA molecular beacons (MBs), introduce a bistable locking mechanism: a closed hairpin conformation that transitions into an extended duplex\supercite{tyagi1996molecular} upon target hybridization, enforcing two distinct, highly predictable emitter separations.~\supercite{bonnet1999thermodynamic,mathur2021understanding,steinhauer2009dna}
\begin{figure} [t!]
  \centering
  \includegraphics[width=\linewidth]{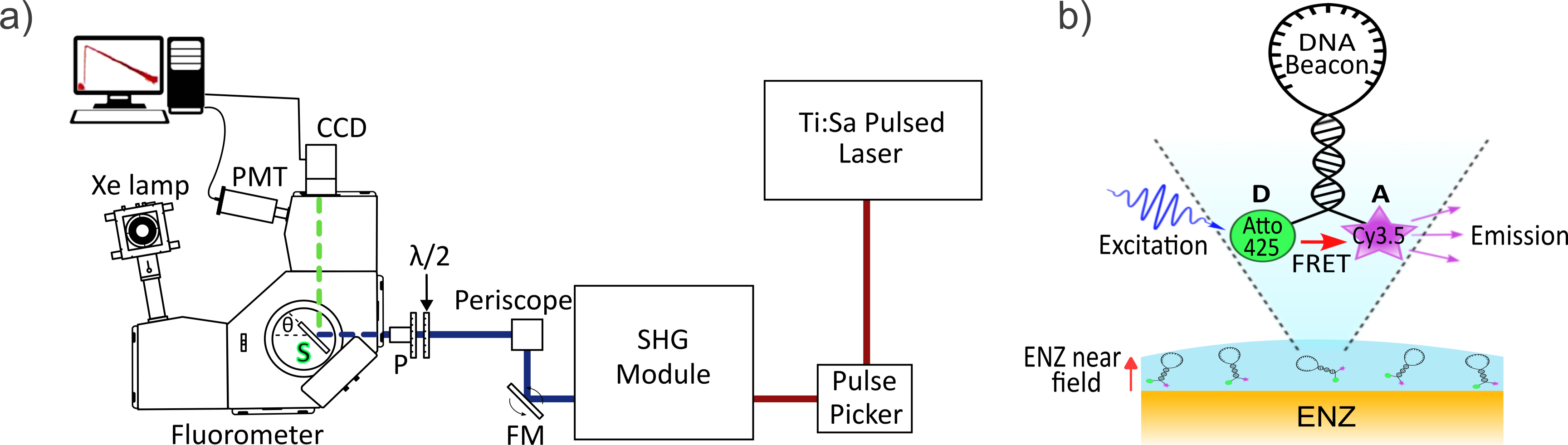}
  \caption{Experimental implementation and system description: a) Experimental layout for steady-state and time-resolved spectroscopy. The sample position denoted by S, is configured for thin film measurements b) Schematic of DNA beacon-based spatial locking of the donor-acceptor pair. Dye-labeled MBs are embedded in a thin PVA film on the ENZ multilayer, positioning fixed-distance FRET pairs within the ENZ near field.}
  \label{fig:setup}
\end{figure}
Here, we unify deterministic DNA emitter placement with ENZ field engineering to enhance Förster resonance energy transfer (FRET) in the visible spectrum beyond singular photonic or structural limits.
%While ENZ effects have been extensively explored in the infrared using low-carrier density semiconductors\supercite{niu2018epsilon}, achieving a comparable response in the visible regime requires sophisticated metamaterial engineering to overcome the inherent loss and dispersion of noble metals\supercite{maas2013experimental}. 
Using DNA MBs, donor-acceptor pairs (Atto425/Cy3.5) are structurally locked into closed hairpin and extended conformations within the near-field of a multilayer Au/TiO$_2$ ENZ platform, tuned to the emitter overlap  band. Steady-state fluorescence and time-resolved photoluminescence (TRPL) measurements quantify the combined effects of both DNA-positioning and ENZ LDOS enhancement (\textbf{Figure~\ref{fig:setup}}), while macroscopic Green tensor modeling resolves the spatial and angular coupling dependencies near the interface. This approach demonstrates a scalable pathway to mitigate nanoscale spatial disorder in bio-sensing and quantum photonics.

\begin{figure}[!ht]
  \centering
  \includegraphics[width=\linewidth]{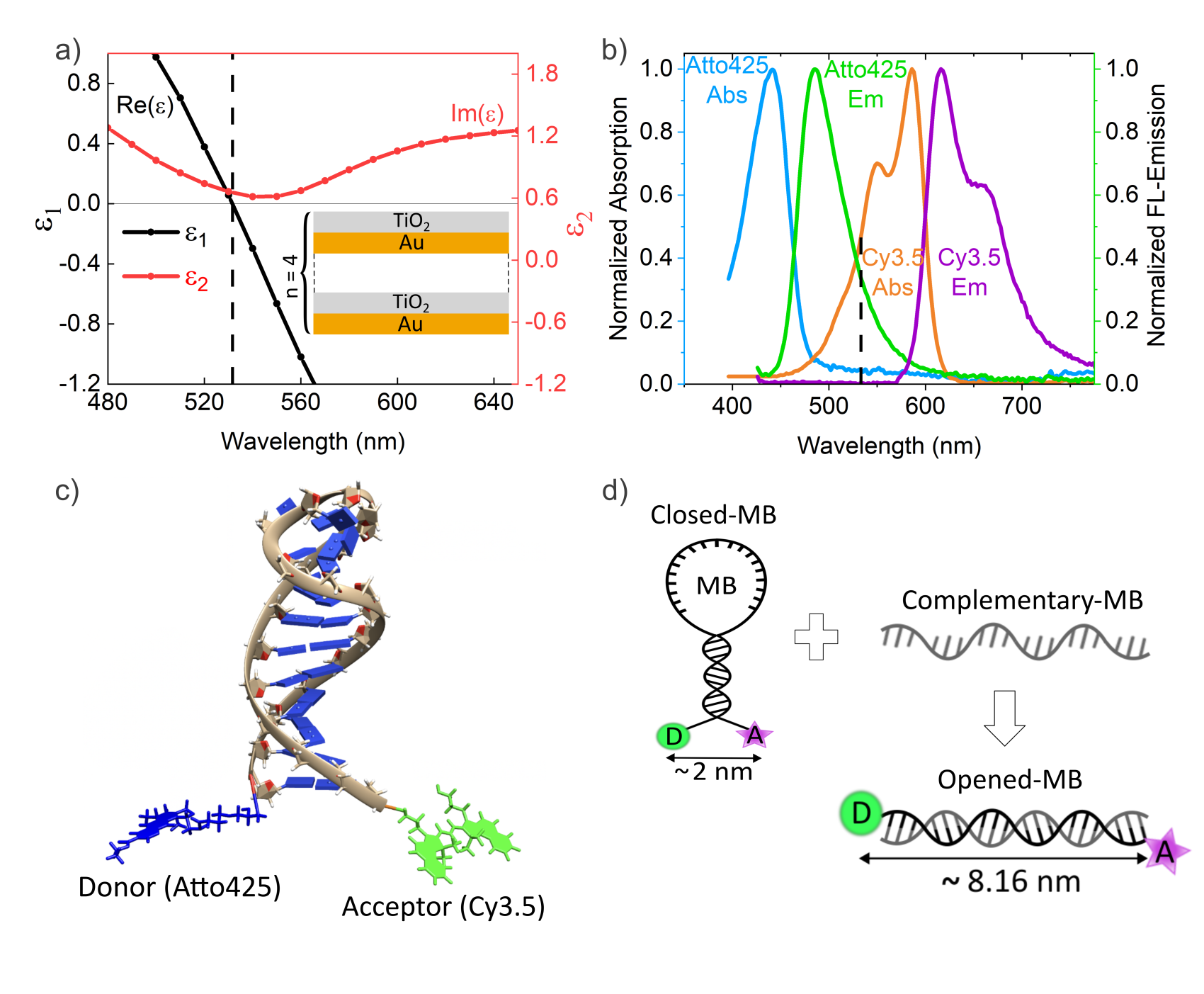}  
  \caption{ENZ and fluorophore selection: (a) Real ($\epsilon_1$, black) and imaginary ($\epsilon_2$, red) parts of the effective permittivity extracted via point-by-point numerical ($n,k$) inversion of the raw spectroscopic ellipsometry data.The vertical dashed line highlights the epsilon-near-zero (ENZ) wavelength where $\epsilon_1=0$. Inset: Schematic cross-section of the Au/TiO$_2$ multilayer ENZ platform ($n$= 4 periods),terminating with a top $\sim$12 nm TiO$_2$ dielectric capping layer in-contact with the polymer superstrate.  (b) Normalized donor emission and acceptor absorption spectra showing the donor-acceptor spectral overlap region. (c) Atomic model of the MB with conjugated donor and acceptor dyes \supercite{abramson2024accurate} (closed MB configuration). (d) Formation of Opened-MB by chemically treating with Complementary-MB to the Closed-MB.}
  \label{fig:2}
\end{figure}

To realize this hybrid coupling, in this study the ENZ platform is engineered as a metal-dielectric multilayer whose zero permittivity crossing spectrally overlaps with the donor acceptor FRET band (\textbf{Figure~\ref{fig:2}a}). The stack consists of four periods of alternating Au (10 nm) and TiO$_2$ (12 nm) deposited via magnetron sputtering with optimized oxygen stoichiometry \supercite{beliaev2021photoluminescence} and adhesion treatments\supercite{sukham2017high}( \textit{Supporting Information (SI) section S2}). While this sub-wavelength stack synthesizes a homogenized ENZ effective medium, near-field layer discretization effects are still important at the microscopic scale (see the \textit{SI, Section S8.7}). The platform terminates with a $\sim$12 nm TiO$_2$ layer, which suppresses direct metallic quenching relative to gold-terminated substrates (\textit{SI, Section S8.6}). Spectroscopic ellipsometry confirmed that the real part of the effective permittivity crosses zero ($\mathrm{Re}(\epsilon) = 0$) at 532 nm, with low dissipative loss ($\mathrm{Im}(\epsilon)$). Atomic Force Microscopy verified sub-nanometer surface roughness ($R_a \sim 0.7\text{-}0.9$ nm), minimizing morphological scattering. \supercite{malureanu2015ultra}

Spectral characterization confirmed the suitability of the Atto425/Cy3.5 pair within this platform, as shown in \textbf{Figure~\ref{fig:2}b} and  \textit{section S3 of the SI}. Donor absorption and emission peak at 439 nm and 488 nm respectively, while Cy3.5 absorbs at 550 nm and emits at 610 nm. Aligning the 532 nm ENZ resonance with this donor-acceptor overlap window selectively tailors the local density of optical states, providing a deterministic platform to probe ENZ-mediated dipole-dipole coupling at extended separations.

Fluorophores were structurally locked using a 24-nucleotide DNA MB conjugated to an Atto425 donor at the 3$^\prime$ end (NHS ester chemistry) and Cy3.5 acceptor at the 5$^\prime$ end (phosphoramidite chemistry) (\textbf{Figure~\ref{fig:2}c} and ~\ref{tab:oligo_si}). The architecture yields two different conformations: a self-hybridized hairpin ("Closed-MB", donor-acceptor separation $\sim 2$ nm\supercite{watson1953molecular}), and an extended duplex formed upon hybridization with a 5-fold molar excess of complementary strand ("Opened-MB", nominal separation $\sim 8.16$ nm, the full separation distribution is shown in \textit{SI, Section S7.3}\}, \textbf{Figure~\ref{fig:2}d} and {Table~\ref{tab:oligo_si}}). This "locking" mechanism significantly reduces geometric ambiguity, allowing changes in quenching or lifetime to be attributed directly to the electromagnetic environment.

To ensure structural stability and uniform dispersion, MBs were embedded in a thin polyvinyl alcohol (PVA) matrix containing 25 mM NaCl, spin coated onto substrates, and cured at 60$^\circ$C (deposition protocol and film thickness in the\textit{SI, Section S1.4,S2.3,} and {Table~\ref{tab:S1}}). This positions the structurally locked donor-acceptor pairs directly within the localized near field of the ENZ substrate (\textbf{Figure~\ref{fig:setup}b}).Optical characterization combined steady-state fluorescence and time-resolved photoluminescence spectroscopy (\textbf{Figure~\ref{fig:setup}a}). Steady state donor quenching spectra were collected on a customized Edinburgh fluorometer using Xenon lamp excitation (320-420 nm, minimizing direct acceptor absorption) and a CCD detector. Picosecond  decay dynamics were resolved via time-correlated single-photon counting (TCSPC)excited by a pulsed Ti:Sapphire laser (average power of 3 W with 80 MHz repetition rate). Pulses were regulated by a pulsed picker and routed through a Second-Harmonic Generation (SHG) module, yielding an on sample excitation power density of 1.05 W/cm$^2$. Donor emission was collected at $\theta \sim 45^\circ$ through a 455 nm long pass filter, onto a photomultiplier tube (PMT) with an instrument response function (IRF) of 5 ps.

Enzymic degradation assay with DNase-I first confirmed that the  donor quenching requires the intact, structurally locked geometry rather than non-specific dye aggregation (\textit{SI} Section S4). Steady-state fluorescence confirmed that in the closed-MB architecture, enzymatic DNA scaffold degradation eliminated donor quenching, demonstrating that efficient energy transfer requires the structurally locked conformation. 

\begin{figure}[!ht]
  \centering
  \includegraphics[width=\linewidth]{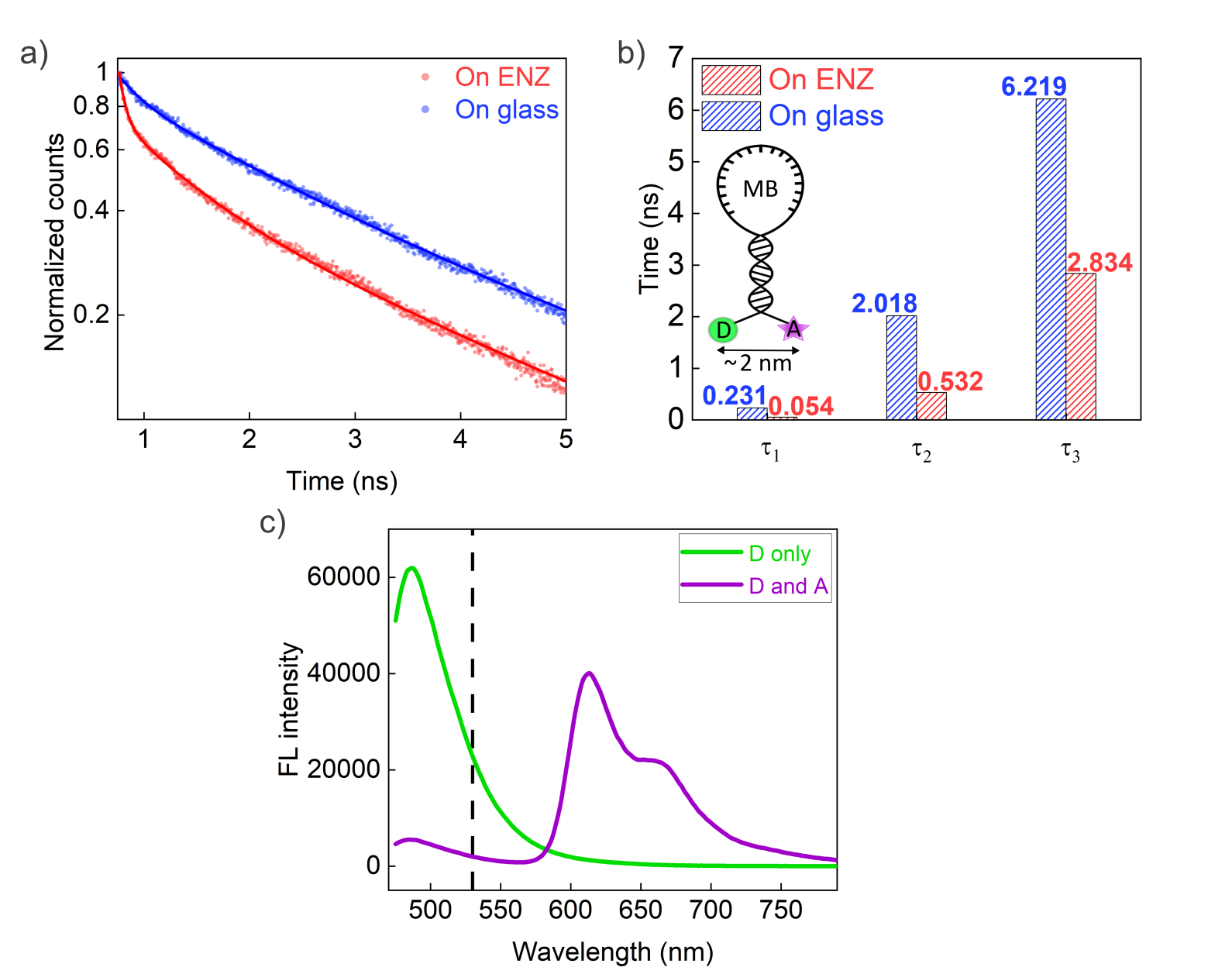}
  \caption{Closed MB system (a) Donor+Acceptor PL decay curves on glass and ENZ substrates (400 nm excitation and 532 nm emission) (b) Histogram showing the fastened decay on ENZ (c) Steady-state fluorescence emission intensity (400 nm excitation) for the closed donor-MB and for the closed MB systems.}
  \label{fig:3}
\end{figure}

In the \textbf{closed-MB} architecture, efficient energy transfer is expected due to the minimal donor-acceptor separation. Two sets of samples were prepared, namely, a control DNA MB functionalized with only the donor fluorophore (Donor-MB) and the target DNA MB functionalized with both the donor and acceptor fluorophores at the 3$^\prime$ and 5$^\prime$ ends, respectively (see {Table~\ref{tab:oligo_si} in \textit{SI}} for more details). Samples were embedded in thin polymer films (\textit{SI} Section 2) on both glass and ENZ substrates to enable direct comparison of emission behavior in different photonic environments. As shown in \textbf{Figure~\ref{fig:3}a}, photoluminescence (PL) decay curves for the donor-acceptor pair demonstrate a significantly faster decay rate on the ENZ substrate compared to glass (reference). The fluorescence decay curve for this emitter pair can be de-convoluted into three distinct $\tau$ components at different scales, including a fast, sub-nanosecond decay channel. The bar plot in \textbf{Figure~\ref{fig:3}b} highlights the significantly decreased relaxation time (increased decay rate) on the ENZ substrate in each decay channel, relative to the glass substrate. This enhancement is a direct measurement of the modified LDOS near the ENZ surface, which strengthens near-field coupling even at small separation distances. These findings were further corroborated by steady-state fluorescence measurements of the donor-only and donor-acceptor closed MBs when excited at the same wavelength (400 nm). Steady-state spectra under 400 nm excitation confirm substantial donor quenching centered at $\sim$ 500 nm accompanied by sensitized Cy3.5 acceptor emission (\textbf{Figure~\ref{fig:3}c}). The energy transfer efficiency calculated \textit{via} spectral quenching: $E = 1 - I_{DA}/I_{D}$, where $I_{D}$ and $I_{DA}$ are donor fluorescence emission intensities (at the donor emission peak wavelength) in the absence and presence of the acceptor, respectively. \supercite{medintz2013fret} This result indicated a $\sim$90\% energy transfer efficiency for the MBs in the solution phase for the closed-MB. Such high efficiency confirmed the reliability of the closed configuration, as it aligns with the expected dipole-dipole coupling at minimal donor-acceptor separations. Quantitative data and the fit model are described in \textit{SI Section S6}.

\begin{figure}[!ht]
  \centering
  \includegraphics[width=\linewidth]{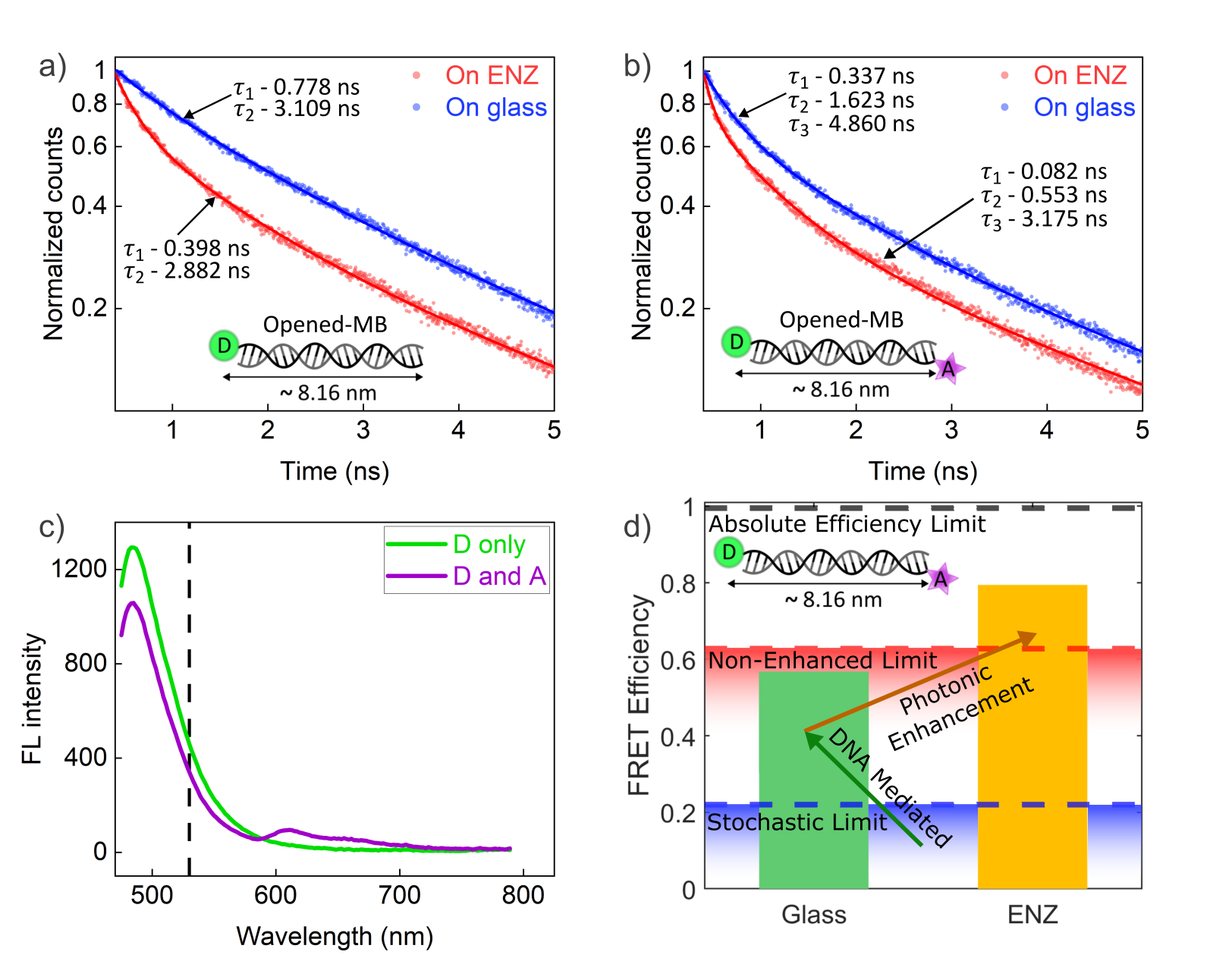}
  \caption{Opened MB system (a) Time-resolved photoluminescence (PL) decay curves for the unfolded beacon containing only the donor dye, measured on a thin polymer film on glass and on the ENZ substrate. (b) PL decay curves for the unfolded beacon containing both donor and acceptor dyes under the same film and substrate conditions (c) Steady-state FL emission opened MB configurations (d) Hierarchical enhancement of energy transfer efficiency. The transition from the stochastic limit (blue dashed line) to the non-enhanced limit on glass (green bar) demonstrates the structural precision provided by DNA-locked architectures, which overcomes the geometric uncertainty typical of random fluorophore dispersions. The subsequent transition from glass to the ENZ substrate (yellow bar) represents the photonic enhancement phase. The efficiency limits are calculated based on the theoretical FRET efficiency at different emitter dipole orientations, including the isotropic average ('Stochastic Limit', $\kappa^2 = 2/3$), and the optimal co-linear configuration ('Non-Enhanced Limit', $\kappa^2 = 4$).}
  \label{fig:4}
\end{figure}

To probe the effect of ENZ photonic enhancement in mediating FRET interactions at increased separation, the \textbf{opened-MB} configuration was tested (formed by hybridizing the MB with its complementary oligo, see Table~\ref{tab:oligo_si} and the \textit{SI, Section S5}). In this design, the estimated donor-acceptor separation is approximately 8.16 nm, corresponding to the theoretical contour length of the 24-base duplex. \supercite{watson1953molecular} Despite the reduced intrinsic coupling at this distance, a pronounced substrate dependence was observed in both the donor only (\textbf{Figure~\ref{fig:4}a}) and donor-acceptor (\textbf{Figure~\ref{fig:4}b}) PL decay curves. 
 For the donor-only opened-MB, the PL decays in glass \textit{versus} ENZ substrate were well described by two components, consistent with the donor’s intrinsic relaxation channels, with both components being shorter on ENZ. This acceleration directly reflects the modification of the LDOS induced by the metamaterial substrate in the absence of the energy transfer.\supercite{purcell1995spontaneous} For the sample with both dyes, the decays are best explained by three components. To separate the substrate-induced LDOS acceleration of the donor from genuine donor-acceptor energy transfer, we differentiate the intrinsic donor decay ($k_{\text{D}} = 1/\tau_{\text{D}}$) from the couples donor-acceptor rate ($k_{\text{D}} = 1/\tau_{\text{DA}}$) via $k_{\text{FRET}} = k_{\text{DA}} - k_{\text{D}}$ (\textit{SI, Section S6.3}). The appearance of an additional component suggests the presence of an additional relaxation pathway associated with donor-acceptor interaction under modified photonic conditions; however, further analysis is required to fully assign its physical origin. Quantitative analysis of the decay times is further discussed in  \textit{Section S6 of the SI}.

The opened-MB system exhibited a FRET efficiency of 56.7\% on glass and 79.4\% on the ENZ substrate according to the lifetime measurements. The FRET efficiency was calculated as $E = 1 - \tau_{DA}/\tau_{D}$, where $\tau_{D}$ and $\tau_{DA}$ are donor lifetimes in the absence and presence of the acceptor, respectively.\supercite{medintz2013fret} The observed efficiency on the glass substrate is substantially larger than the theoretical $\sim$21\% value predicted for the emitter pair in solution without photonic enhancement and with stochastic dipole alignment. This discrepancy can be understood quantitatively within the Förster framework. The fundamental equation for efficiency as a function of radius $r$ is: 
\begin{equation}
    E = (1 + (r/R_0)^6)^{-1},
\end{equation}
where the Förster radius $R_0$ is given by:
\begin{equation}
    R_0^6\, \propto\, \kappa^2\, Q_D\, J\, n^{-4},
\end{equation}
for the orientation factor $\kappa^2$, the donor quantum yield $Q_D$,the spectral overlap integral $J$, and the local refractive index $n$. \supercite{lakowicz2006principles}
The quoted value for the theoretical stochastic efficiency (21\%) corresponds to a Förster radius of $R_{0,theory} \sim 5.77$ nm, calculated from standard solution parameters.~\supercite{wu1994resonance}
%while the observed efficiency on glass implies an effective $R_{0,eff} \sim 7.53$ nm ($\sim$1.3$\times$ larger). 
The observed difference in the experimentally measured efficiency can be interpreted as changes in the effective Förster radius, which can be influenced by each of these factors. Of particular importance for DNA-mediated emitter structures, the emitter pairs in solution assume a random orientation with $\kappa^2 = 2/3$. However, this is below the physical maximum of 4 and generally not true in DNA structures where the dipole orientation is physically restricted from probing all random orientations.\supercite{mathur2021understanding} In practice, several thin-film-related effects may contribute simultaneously, including: i) restricted rotational freedom and partial dipole alignment due to both the DNA and PVA film (increasing $\kappa^2$); ii) changes in donor quantum yield $Q_D$ due to suppressed non-radiative channels in the solid matrix; iii) local differences in refractive index compared to aqueous solution; and iv) interface-induced ordering near the glass substrate. A more comprehensive theoretical analysis of transfer coupling enhancement with respect to dipole position and orientation is provided in the \textit{SI}. These effects could conceivably increase $R_0$ and hence the observed FRET efficiency at longer distances. In summary, a consistent explanation for the increased efficiency on glass is orientational and photo-physical modifications introduced by the thin-film environment.\supercite{biehs2016long,schneckenburger2020forster,jones2019resonance} Solution phase steady-state fluorescence benchmarking confirmed the 21\%  baseline (\textbf{Figure~\ref{fig:4}c}), establishing that the elevated glass efficiency is driven by matrix-induced ordering and restricted dye rotation.

The findings of this study’s contribution to nanoscale energy transport is summarized in \textbf{Figure~\ref{fig:4}d}. We identified two distinct regimes of enhancement that allow the system to surpass conventional energy transfer boundaries. First, the transition from the stochastic limit, characterized by the broad uncontrolled donor-acceptor distributions found in traditional colloidal systems, to the DNA-mediated enhancement on glass demonstrates the power of the molecular locking mechanism.\supercite{akselrod2014visualization} By constraining the emitters to a fixed 8.16 nm separation within the rigid DNA framework, we establish a reproducible FRET efficiency (56.7\%) that already represents a significant optimization over stochastic assembly.Upon placing these locked emitters on the ENZ platform, a second photonic enhancement leap occurs, driving the FRET efficiency to 79.4 \% and surpassing the non-enhanced glass limit via tailored LDOS boundary conditions. 

To quantify the impact of the ENZ environment on energy transfer, we associate the shortest lifetime component ($\tau_1$) with the interacting donor fraction undergoing active energy transfer. In heterogeneous systems, a fraction of the donor population often remains non-interacting due to stoichiometric incompleteness or orientation effects, resulting in a multi-exponential decay profile.~\supercite{BeckerHicklFRET} By utilizing a multi-exponential model and isolating the $\tau_1$ channel, we isolate an apparent transfer efficiency representative of the interactive donor population, mitigating the averaging effects of non-interacting emitters. In our DNA-locked architecture, the donor system is accurately described by a double-exponential fit, accounting for the primary emitter decay and minor environmental fluctuations. However, for the donor-acceptor (Opened-MB) system, a triple-exponential model is required to maintain a $\chi^2$ close to unity and minimize residual patterns. This third component likely reflects the increased complexity of the coupled system within the ENZ-Purcell regime. \supercite{akselrod2014probing} Regardless of the additional components, the introduction of the acceptor consistently manifests itself as a pronounced acceleration of the specific $\tau_1$ channel.
On the glass reference, the introduction of the acceptor reduced the donor's $\tau_1$ from 0.778 ns to 0.337 ns, while the FRET-active component reached 0.082 ns corresponding to an enhanced efficiency of 79.4\%. Beyond efficiency, the FRET-associated transfer rate ($k_{ET}$)extracted from the $\tau_1$ values provides a measure of coupling acceleration. On glass, the transfer rate is approximately 1.68 ns$^{-1}$ whereas on the ENZ substrate it accelerates to 9.68 ns$^{-1}$, a 5.8 fold increase in the fast transfer rate. This acceleration is significant for the opened configuration, where conventional Förster coupling is actually weak. 

\begin{figure}[h!]
    \centering
    \includegraphics[width=\textwidth]{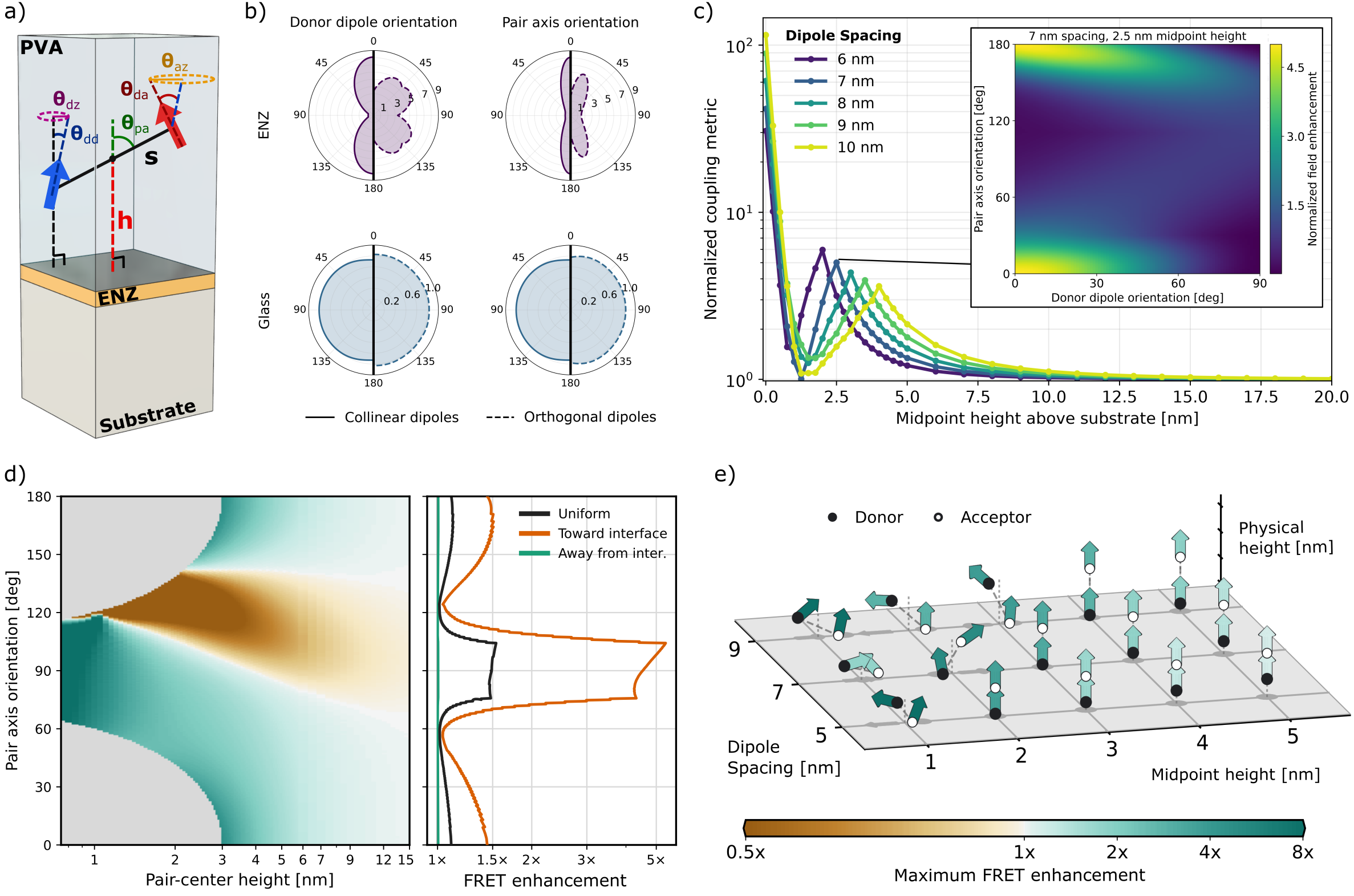} 
    \caption{Theoretical simulation of substrate-mediated donor-acceptor coupling. 
    \textbf{a} Schematic of the simulation geometry. A donor and acceptor dipole pair is embedded in PVA above an ENZ film on a glass substrate. The emitter pair is defined by separation $s$, midpoint height $h$, pair-axis orientation $\theta_{\text{pa}}$, and donor / acceptor dipole orientations $\theta_{\text{dd}}$, $\theta_{\text{dz}}$, $\theta_{\text{az}}$, and $\theta_{\text{da}}$. 
    \textbf{b} Angular dependence of the normalized coupling enhancement with explicit multilayer ENZ (top row) and glass (bottom row) substrates for the emitter pair at \qty{8}{\nm} dipole spacing and a fixed midpoint at \qty{3}{\nm} above the interface. Relative dipole orientations are shown with solid (collinear) and dashed (orthogonal) lines, respectively. The shaded regions demonstrate the range of enhancement values considering all open degrees of freedom.
    \textbf{c} Maximum normalized transfer enhancement versus donor-acceptor midpoint height above the substrate for dipole spacings from \qtyrange{6}{10}{\nm} for an explicit multilayer ENZ substrate. The donor-acceptor pair show strong enhancement near the substrate, mainly contained within $h < \qty{10}{\nm}$ and decaying toward unity with increasing height. \textbf{(inset)} Orientation dependence of the normalized coupling enhancement with $\theta_{\text{pa}}$ and $\theta_{\text{dd}}$, at \qty{7}{\nm} spacing and \qty{2.5}{\nm} midpoint height above the ENZ substrate. The magnitude of the transfer efficiency enhancement is strongly dependent upon the orientation of the dipoles.
    \textbf{d} Maximum FRET transfer efficiency enhancement for the dipole pair above the multilayer ENZ over glass substrates with 8~nm donor-acceptor spacing. \textbf{(left)} The maximal enhancement factor over all non-specified orientations as a function of $\theta_{\text{pa}}$ and h (log spaced). Gray regions represent disallowed orientations. \textbf{(right)} Height-averaged FRET enhancement profiles using three spatial averaging techniques. Note that this represents an upper bound rather than a true orientational average.
    \textbf{e} Optimal dipole orientation configurations (maximizing over all non-specified orientations) for maximally enhanced FRET transfer-rate enhancement with fixed dipole spacing and pair midpoint height (h). Donor and acceptor dipoles are represented as arrows with projection shadows on the substrate plane. FRET enhancement can be observed for at least one configuration at each height and spacing point considered. 
    }
    \label{fig:fig5}
\end{figure}

While $\tau_1$ provides the most direct probe of the interacting population, we also performed an ensemble-level analysis using the amplitude-weighted average lifetimes ($\langle\tau\rangle$). This provides a complementary perspective consistent with the established literature for characterizing heterogeneous FRET systems.\supercite{kawachiya2019photoluminescence,spriet2008enhanced} These results and their comparison to the component-specific analysis are detailed in the Section S6.2 of the \textit{SI}. This ensemble averaged parameter represents the global intensity-equivalent decay of the entire donor ensemble (showing a rate increase from 0.448 ns$^{-1}$ on glass to 0.705 ns$^{-1}$ on ENZ). The distinction between $\tau_1$ and $\langle\tau\rangle$ is corroborated by the heterogeneous emitter population observed in our simulations (\textbf{Figure~\ref{fig:fig5}} and Figures S9 - S11), where substrate-mediated enhancement is strongly confined near the interface and varies from suppression ($<0.5\times$) to $>5\times$ enhancement with configuration. $\tau_1$ can be associated with the most enhanced sub-population, while $\langle\tau\rangle$ averages across the 34 nm film, yielding a $1.57\times$ ensemble enhancement consistent with uniform height-averaged predictions (\textbf{Figure~\ref{fig:fig5}d}). However, we note that fully distinguishing FRET-specific enhancement from general photonic contributions in these complex media may require further characterization, such as transient absorption spectroscopy (TAS), to independently resolve stimulated emission dynamic and excited-state populations.

To determine whether the experimentally observed decay rate acceleration is consistent with donor-acceptor coupling mediated by the ENZ substrate, the opened-MB DNA system is modeled as two classical point dipoles embedded in PVA above the measured multilayer ENZ structure; see \textbf{Figure~\ref{fig:fig5}a}. The DNA is represented implicitly through the donor-acceptor separation, also constraining the allowed pair orientation geometries. The electromagnetic Green tensor is written as the sum of homogeneous-PVA and substrate-reflected contributions, $\mathbf{G}=\mathbf{G}_{0}+\mathbf{G}_{\mathrm{ref}}$. The results are presented both as FRET rate enhancement ratios for the ENZ substrate with respect to a homogeneous PVA reference --- to avoid large enhancements due to unfavorable initial configurations --- and for the same dipole configurations over a glass substrate --- which is the experimentally relevant enhancement metric. A range of spacings are simulated to cover the distribution of possible experimental configurations (\textit{SI}, Section S7.3). Full numerical details and the donor-only LDOS analysis are provided in the \textit{SI}.

The normalized coupling metric (FRET transfer efficiency proxy) is strongly anisotropic near the ENZ interface. The range of achievable coupling metrics at different dipole orientations for a range of donor dipole and pair axis angles is shown in \textbf{Figure~\ref{fig:fig5}b}. Changing the dipole configuration can produce both enhanced and suppressed responses through interference between the direct and reflected donor fields. The angular dependence is substantially weaker for the isotropic glass substrate. This indicates that different dipole configurations above the same substrate can, in principle, produce both highly enhanced and quenched FRET responses. The interaction is also highly spatially confined. As shown in \textbf{Figure~\ref{fig:fig5}c}, the largest ENZ-induced enhancement occurs within the first several nanometers of the interface and approaches the homogeneous response (unity) beyond approximately 10~nm, implying that a significant volume of a $\sim$30~nm thick PVA film would experience minimal enhancement from the ENZ substrate. Also note that the position of the off-substrate maximum changes with donor-acceptor separation distance because the pair orientation is geometrically restricted to dyes physically above the substrate.

\textbf{Figure~\ref{fig:fig5}d} compares the transfer-rate enhancement directly between the explicit ENZ multilayer and glass for an 8~nm donor-acceptor separation. The colormap fixes the donor polar orientation at 0$^{\circ}$ from the surface normal and reports the maximum enhancement over all other dipole orientation configurations. Note that this represents an upper bound on the FRET transfer-rate enhancement for the optimal dipole configurations, rather than an orientational average. For different pair-axis orientations at the same distance from the substrate, it is possible to achieve a range of enhancements from quenching ($<1\times$) to $>5\times$ enhancement. The adjacent profiles show the maximal FRET transfer-rate enhancement ratio over all non-specified orientation configurations at each height averaged over an entire 34~nm PVA film. Three height distributions are considered: uniform, exponentially weighted toward the substrate interface (with a 5~nm decay length), and exponentially weighted away from the substrate interface with the same decay length biased toward the air-PVA interface. For 8~nm dipole separation, closely representing the experimental 8.16~nm opened-MB, the interface-weighted height-averaged profile reaches $>5\times$ enhancement for favorable pair-axis orientations. This is qualitatively similar to the experimentally extracted $\tau_1$ enhancement of approximately $5.8\times$. In contrast, the uniform and away-from-interface profiles reach approximately $1.5\times$ and $1\times$, respectively. The magnitude of the FRET enhancement also increases with increasing donor-acceptor separation, where the unmodified transfer interaction is weaker, as discussed in the \textit{SI}, indicating that wider separations could reach even greater enhancement factors.

Altogether, these simulations support the interpretation that a range of photonic FRET transfer-rate enhancements can be present for different sub-populations of a heterogeneous ensemble of dipole pairs. In optimal configurations, near-interface sub-populations can experience an ENZ-modified enhancement comparable to the experimentally observed fast $\tau_1$ channel, while dipole pairs at greater heights or exhibiting suboptimal configurations can contribute a response close to unity and dilute the ensemble-averaged enhancement. Note that the model agreement should be taken primarily as qualitative due to several limitations, including monochromaticity, unknown empirical linker and DNA conformational distributions, and unaccounted photon-count bias. Future studies specifying or constraining the range of orientational configurations could further improve the experimental correspondence. However, the theoretically obtained enhancement factors are plausibly consistent with the observed experimental enhancement magnitudes and heterogeneous enhancement over different independent subpopulations. 

Finally, \textbf{Figure~\ref{fig:fig5}e} shows the optimized dipole configurations for maximal FRET transfer enhancement for ENZ and glass substrates over a range of donor-acceptor spacing and pair midpoint heights above the substrate. The donor and acceptor positions are shown as circles, and the arrows represent the optimal dipole orientations, with color showing the magnitude of the FRET transfer-rate enhancement. A broader analysis of these configurations is given in the \textit{SI}. For dipole pairs far from the substrate interface, the optimal configuration converges toward both dipoles being collinear with a pair axis normal to the interface ($\kappa^2=4$). However, due to the geometrical confinement in the ENZ near-field, a high degree of enhancement can also be obtained for non-collinear dipole arrangements at tilted pair axes close to the substrate interface. Therefore, obtaining maximized coupling depends heavily on controlling the equilibrium orientation distribution. This analysis could be useful when dipole configurations can be experimentally controlled, enabling precise placement of dipole states in uniformly enhanced configurations.

 In conclusion this study experimentally demonstrates that ENZ environments can significantly enhance dipole-dipole coupling between fluorophores at molecular scales. Using DNA molecular beacons to establish structurally locked as nanoscale scaffolds for fixed donor acceptor separations, we observed accelerated donor decay and enhanced FRET efficiency for both closed and opened configurations when the system was placed on a multilayer ENZ substrate. These results provide direct evidence that ENZ boundary conditions strongly modify near-field donor-acceptor coupling at molecular distances beyond conventional distance-limited regimes without altering molecular geometry. Theoretical modeling corroborates these observations, showing that the substrate-mediated enhancement is tightly confined within the first few nanometers of the ENZ interface and exhibits pronounced sensitivity to partial separation and dipole orientation. Favorable spatial and angular dipole distributions can yield a greater than five-fold enhancement in the FRET transfer-rate while suboptimal orientations lead to transfer suppression. By leveraging zero-index phase uniformity, this visible ENZ platform resolves conventional  planar confinement-dephasing trade-offs, realizing an external 5.8-fold transfer rate acceleration directly across an accessible interface. . Crucially, our electrodynamic modeling demonstrates that this macroscopic zero-index dispersion operates with pronounced orientational selectivity, selectively amplifying transfer rates for structurally aligned molecular dipoles.  Moving forward, these orientation-sensitive boundary maps provide a direct blueprint for designing higher-order DNA origami sheets and rigidified scaffolds engineered to lock emitter dipoles into the predicted favorable orientations, maximizing near-field energy routing across macroscopic nanophotonic interfaces. Ultimately, combining DNA-programmed structural precision with macroscopic zero-index boundary conditions establishes a deterministic platform for engineering quantum electrodynamic phenomena, opening new avenues for planar light-harvesting systems, bio-sensing, and solid-state quantum photonics.
\section*{Supporting Information}

The Supporting Information is available free of charge at (Link)

Detailed experimental methods for MB preparation and film deposition; ENZ multilayer fabrication protocols; surface and thickness characterization with ellipsometry, and profilometry; fluorophore photophysical characterization; enzymatic DNase-I degradation assay protocols and spectra; opened-MB baseline steady-state measurements; multiexponential TCSPC decay fitting parameters and amplitude-weighted lifetime analysis; theoretical FRET parameters; structural molecular modeling of fluorophore linkers; Green's tensor and FDTD numerical simulation frameworks; orientation-dependent coupling maps; comparison with metallic substrates and ENZ multilayer discretizations.  

\section*{Acknowledgments}

The authors acknowledge the use of the Materials for Optoelectronics Research and Education (MORE) Center, a core facility at Case Western Reserve University (est. 2011 via Ohio Third Frontier grant TECH 09-021). The authors acknowledge support from the Ohio Third Frontier Project “Research Cluster on Surfaces in Advanced Materials” (RC-SAM) at Case Western Reserve University. Additionally, authors acknowledge the financial support from the Graduate Student Scholarship and Creative Endeavors (GSSCE) grant provided by the Graduate Council of Arts and Sciences (GCAS) at Case Western Reserve University. The authors also thank Kayla Neyra and Heather Everson for their assistance with chemical preparations.

% Main Text References appear immediately after the main text
\printbibliography[heading=subbibliography, title={References}]
\end{refsection}
% --- MAIN TEXT ENDS HERE ---
\newpage
\begin{refsection} % Starts independent citation numbering [1, 2, ...] for SI

% Reset and prefix counters for SI
\setcounter{section}{0}
\setcounter{figure}{0}
\setcounter{table}{0}
\setcounter{equation}{0}
\setcounter{page}{1}

\renewcommand{\thesection}{S\arabic{section}}
\renewcommand{\thefigure}{S\arabic{figure}}
\renewcommand{\thetable}{S\arabic{table}}
\renewcommand{\theequation}{S\arabic{equation}}
\renewcommand{\thepage}{S\arabic{page}}

% SI Title and Author Header Block
\begin{center}
  {\Large \textbf{Supporting Information}} \\[0.5em]
  {\LARGE \textbf{Enhancing FRET through DNA-controlled Emitters and ENZ Metamaterials}} \\[1.5em]
  
  {\large Akeshi Aththanayake,$^{1}$ Andrew Lininger,$^{1}$ Khoi-Anh Pham,$^{1}$ Radu Malureanu,$^{2,3}$ Divita Mathur,$^{4,*}$ and Giuseppe Strangi$^{1,5,6,*}$} \\[1em]
  
  \small
  $^{1}$Department of Physics, Case Western Reserve University, Cleveland, OH, USA \\
  $^{2}$Department of Electrical and Photonics Engineering, Technical University of Denmark, Copenhagen, Denmark \\
  $^{3}$National Centre for Nano Fabrication and Characterization, Technical University of Denmark, Copenhagen, Denmark \\
  $^{4}$Department of Chemistry, Case Western Reserve University, Cleveland, OH, USA \\
  $^{5}$Department of Physics, NLHT Lab - University of Calabria and CNR-NANOTEC Istituto di Nanotecnologia, Rende, Italy \\[0.8em]
  $^{6}$Case Comprehensive Cancer Center, Case Western Reserve University, Cleveland, OH, USA \\[0.8em]
  
  $^{*}$Email: divita.mathur@case.edu, giuseppe.strangi@case.edu
\end{center}
\vspace{0.5em}
\section*{S1. Materials and Methods}

\subsection*{S1.1 Chemical Reagents}
Polyvinyl alcohol (\ce{[-CH2CH(OH)-]n}) was purchased from Sigma-Aldrich (Catalog \#363138-25G), sodium chloride (\ce{NaCl}) was purchased from Sigma-Aldrich (Catalog \#S9888-25G), molecular biology grade water was obtained from Cytiva (Catalog \#SH30538.03), phosphate-buffered saline (PBS) solutions were obtained from Cytiva (Catalog \#SH30258.02), EDTA was purchased from FisherBiotech (Electrophoresis grade, catalog \#BP118-500), and DNase~I (RNase-free) was purchased from New England Biolabs (Catalog \#M0303L). All chemicals were used as received.

\subsection*{S1.2 Oligonucleotide Sequences}
Single-stranded DNA oligonucleotides, with and without fluorescent dye labels, were purchased from Integrated DNA Technologies (IDT) in lyophilized form. All oligonucleotides were reconstituted in molecular biology grade water to a stock concentration of \SI{100}{\micro\Molar} and stored at \SI{4}{\degree C} prior to use. There are two predominant molecular beacon (MB) samples used in this study. One is the ``closed'' or formed MB, which comprises a DNA oligonucleotide forming the molecular beacon (MB) conjugated to a donor dye (ATTO~425) at the 3$^\prime$ end and an acceptor dye (Cy3.5) at the 5$^\prime$ end. A complementary strand was also acquired to prepare the ``opened'' MB conformation. Additionally, a donor-only oligo (referred to as ``donor-MB'') was purchased to carry out studies involving the donor only. The sequences and terminal chemical modifications are summarized in  \textbf{Table~\ref{tab:oligo_si}}.
\begin{table}[!ht]
\centering
  \caption{List of oligos and the chemical modifications.}
  \label{tab:oligo_si}
  \begin{tabular}{@{}lll@{}}
    \hline
    \textbf{Name} & \textbf{Sequence} (5$^\prime \rightarrow 3^\prime$) & \textbf{Modification} \\
    \hline
    MB & GAA TTC GGT ATT TCC TCC GAA TTC & \textbf{5$^\prime$} Cy3.5;\quad \textbf{3$^\prime$} ATTO425 \\
    Complementary MB & GAA TTC GGA GGA AAT ACC GAA TTC & N/A \\
    Donor-MB & GAA TTC GGT ATT TCC TCC GAA TTC & \textbf{3$^\prime$} ATTO425 \\
    \hline
  \end{tabular}
\end{table}
\subsection*{S1.3 Preparation of Molecular Beacon-PVA Composite}
To ensure structural stability and a static spatial dispersion of the molecular beacons (MBs) during spectroscopic measurements, the MBs were embedded in a thin polyvinyl alcohol (PVA) film on the ENZ substrate. A \SI{22}{\micro\Molar} solution of the MB (\SI{100}{\micro\L}) was prepared in molecular biology grade water. This was subsequently mixed with a PVA-\ce{NaCl} solution to achieve a final MB concentration of \SI{11.43}{\micro\Molar}. The \ce{NaCl} concentration was adjusted to a final value of \SI{25}{\milli\Molar} in bio-grade water to maintain the ionic strength required for DNA stability. These batches were stored at \SI{4}{\degree C} for a short term prior to film deposition. Throughout this study, the specific mixture containing MBs, PVA, and \ce{NaCl} is referred to as the ``opened/closed MB system.''

\subsection*{S1.4 Spin Coating and Film Preparation}
The thin PVA films were fabricated using the above mixture on cleaned \qtyproduct{0.5 x 0.5}{\cm} borosilicate cover glasses using a Laurell spin coater. A two-step spin-coating process was employed:
\begin{enumerate}
    \item \qty{500}{\rpm} for \qty{5}{\second} to initiate spreading
    \item \qty{5000}{\rpm} for \qty{20}{\second} at an acceleration rate of \qty{1000}{\rpm\per\second} to define the final thickness
\end{enumerate}
Immediately following the film deposition, the samples were hard-baked on a hot plate at \qty{50}{\degree C} for \qty{10}{\minute} to thermally polymerize the PVA film. The thin films on ENZ were fabricated in a similar manner.
\section*{S2. ENZ substrate fabrication and Characterization}

\subsection*{S2.1 Multilayer Deposition Protocol}
The epsilon-near-zero (ENZ) multilayer substrates used in this study were fabricated at the Technical University of Denmark (DTU). The complete characterization of the multilayer structure can be found elsewhere\supercite{Leandro2015}.

The samples were prepared on fused silica substrates following a multistep thin-film deposition procedure designed to achieve an effective permittivity crossing zero in the visible spectral range while minimizing optical losses. Experimental data for this is illustrated in main text figure 2. Prior to metal and oxide deposition, the substrates (coverslips) were functionalized with a self-assembled monolayer of (3-aminopropyl)trimethoxysilane (ATPMS). The APTMS layer was deposited by immersing the substrates in a solution consisting of 2.5\% APTMS, 2.5\% \ce{H2O}, and 95\% isopropyl alcohol (IPA),\supercite{sukham2017high} following established protocols. This surface treatment promotes adhesion and enables controlled multilayer growth.

Following surface functionalization, a thin gold (Au) layer (\qty{10}{\nm}) was deposited by magnetron sputtering~[99.999\% purity] under low-pressure and high-power conditions to ensure the formation of a continuous metallic film with low roughness ($R_a = (0.8 \pm 0.1)\text{ nm}$) measured with atomic force microscopy. After Au deposition, a second APTMS layer was applied using the same solution-based procedure to prepare the surface for subsequent oxide deposition.

Titanium dioxide (\ce{TiO2}) layers (\qty{12}{\nm} each layer) were then deposited by sputtering a titanium target in an oxygen-containing atmosphere, with oxygen concentration approximately 10\% to ensure stoichiometric \ce{TiO2} formation. The sequence of APTMS functionalization, Au deposition, APTMS re-functionalization, and \ce{TiO2} deposition was repeated 4 times to construct a multilayer metal/dielectric stack.The platform terminates with a $\sim$12 nm TiO$_2$ layer, which suppresses direct metallic quenching relative to gold-terminated substrates.

\subsection*{S2.2 Surface Characterization and Thickness Verification}
Film thicknesses were verified using a combination of methods.

\textbf{Stylus Profilometry:} A Tencor P-6 stylus profilometer (force \qty{0.5}{\mg}, \qty{2}{\micro\m} ball tip) scanned step edges to verify opaque film heights (on glass). Scan length [\qty{1023}{\micro\m}] and scan speed [\qty{100}{\micro\m\per\s}] were kept constant for all measurements. Four line scans were acquired at four different positions on each sample on glass, and the film thickness was determined by averaging the measured step heights.

\textbf{Optical profilometry:} To corroborate the stylus measurements, a Zygo NewView 7300 Scanning White Light interferometry (SWLI) system was employed. Surface morphology analysis confirmed that the fabricated films exhibit high spatial uniformity across the central region of the substrate. Crucially, the film thickness remains roughly constant within the illumination area of the excitation laser, ensuring that the probed molecular beacons experience a consistent local environment and offset from the substrate. Thickness values were extracted by analyzing height difference across a well-defined step edge between coated and uncoated regions.

\textbf{Spectroscopic Ellipsometry:} A J.A. Woollam V-VASE system was used to confirm the effective permittivity zero-crossing (ENZ point) in the ENZ multilayer substrates at \qty{532}{\nm}. Additionally, ellipsometric data were analyzed using an oscillator-based optical model to confirm the thickness and permittivity of the film deposited on ENZ.

\begin{table}[h!]
\centering
\caption{Average film thickness for each system obtained from above techniques}
\label{tab:S1}
\begin{tabular}{lcc}
\toprule
System & \begin{tabular}[c]{@{}c@{}}Average thickness from\\ profilometry (nm)\end{tabular} & \begin{tabular}[c]{@{}c@{}}Thickness obtained from\\ ellipsometry (nm)\end{tabular} \\
\midrule
Closed-MB on glass & $31.53 \pm 0.21$ & -- \\
Closed-MB on ENZ & -- & $41.56 \pm 0.676$ \\
Opened donor-MB on glass & $27.82 \pm 0.38$ & -- \\
Opened donor-MB on ENZ & -- & $34.79 \pm 0.344$ \\
Opened-MB on glass & $26.77 \pm 0.28$ & -- \\
Opened-MB on ENZ & -- & $33.85 \pm 2.96$ \\
\bottomrule
\end{tabular}
\end{table}

\noindent While identical spin-coating parameters were maintained across all samples, the observed variations in film thickness between glass and ENZ substrates are attributed to differences in surface energy and the resulting wettability of the aqueous PVA/MB solution on the respective surface.

\subsection*{S2.3 Film thickness considerations}

The choice of a $\sim30$~nm polyvinyl alcohol (PVA) film thickness was guided by a careful balance between biological stability, film morphology, and spectroscopic signal-to-noise ratio:

\paragraph*{Film uniformity and morphological integrity:} Reducing the PVA film thickness below 10~nm poses severe fabrication hurdles. At sub-10~nm regimes, spin cast polymer films are highly susceptible to film rupture and island formation during thermal curing, particularly over the underlying oxide surface. In our preliminary optimization, we observed that extremely thin films compromise optical uniformity and lead to localized pinholes and discontinuous domains, which complicate the optical environment and interpretation.

\paragraph*{DNA dimensional constraints and conformation:} The rigid 24-base-pair opened duplex possesses a physical contour length of 8.16 nm, while the terminal flexible linkers add an effective footprint of $\sim 1.16$~nm (3’-Atto425) and $\sim 0.73$~nm (5’-Cy3.5) yielding a total molecular dimension of approximately 10~nm. In a passive spin coating matrix without specific surface tethering, confining the solution into a sub-10 nm slab can induce shear stress, steric hindrance, and rapid solvent evaporation, risking perturbation to the duplex hybridization or non-specific dye-surface adsorption.

\section*{S3. Optical Characterization of Fluorophores}

\subsection*{S3.1 Absorbance Measurements}
Donor (Atto425) and acceptor (Cy3.5) dyes were selected for high spectral overlap and compatibility with the available Ti:sapphire laser excitation range (\qtyrange{400}{420}{\nm}). The absorption spectra of the fluorophores conjugated to the DNA scaffolds were measured using a Cary series UV-VIS spectrometer [Agilent Technologies]. Samples (\qty{400}{\micro\L}) of the closed MB systems were loaded into quartz cuvettes with a fixed optical path length. UV-VIS absorbance spectra were recorded over a wavelength range of \qtyrange{350}{750}{\nm}. Molecular biology grade water mixed with PVA solvent was utilized as the blank. All measurements were performed at room temperature and repeated twice to ensure the reproducibility of the steady-state curves.

\subsection*{S3.2 Fluorescence Spectroscopy}
Steady-state fluorescence spectra for the donor-MB and donor-acceptor MB systems were acquired using a custom-built Edinburgh Instruments spectroscopic fluorometry setup. To capture the emission profiles, samples were excited at \qty{420}{\nm} with a \qty{10}{\nm} bandwidth, and emission was collected from \qtyrange{475}{800}{\nm}.

\subsection*{S3.3 Optical Characterization Results}
\begin{figure}[h!]
    \centering
    \includegraphics[width=\textwidth]{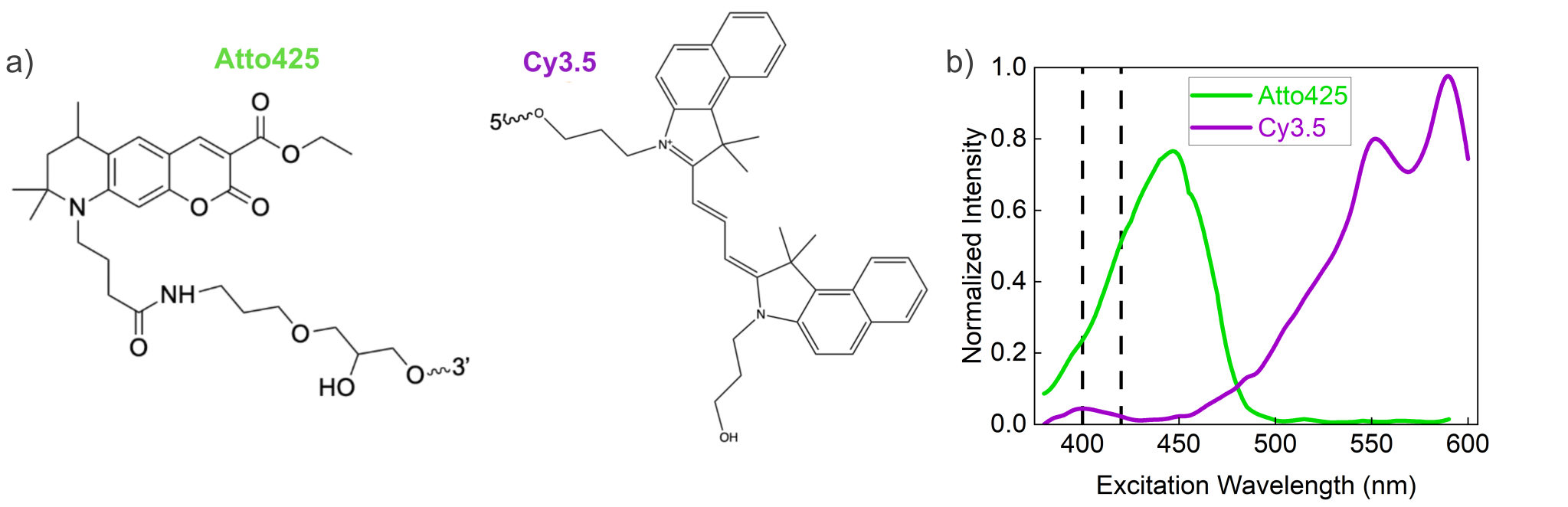} 
    \caption{Molecular structural analysis a) Chemical structures of Atto 425 and Cy3.5 produced by ChemDraw 3D b) Fluorescent emission profiles under varying excitation wavelengths.}
    \label{fig:S1}
\end{figure}

The resulting absorption and emission spectra, obtained through fluorescence emission spectroscopy, are shown in Figure 2b) in the main text. The absorption peaks observed in the spectrum align well with the literature values for the respective dyes. Atto 425, the donor dye exhibited an absorption peak at \qty{439}{\nm}, while Cyanine 3.5, the acceptor dye, showed an absorption peak at \qty{550}{\nm}. Similarly, for the emission, Atto 425 displayed a peak at \qty{488}{\nm}, and Cy3.5 showed a peak at \qty{610}{\nm}. The absence of any other significant absorption or emission peaks in the spectrum confirms the successful incorporation of the dyes into the DNA oligonucleotide and their expected optical properties. \textbf{Figure~\ref{fig:S1}a} shows the chemical structures of the two fluorophores used in this study. To understand the excitation wavelength of the system, an excitation series is done from \qtyrange{300}{500}{\nm} to observe the emission spectrum as illustrated in \textbf{Figure~\ref{fig:S1}b}. It is clear that in the range of \qtyrange{400}{420}{\nm} the direct excitation of the acceptor is very low.

\section*{S4. FRET Validation: Enzymatic Degradation Assay}

\subsection*{S4.1 Experimental Protocol}
Given the high predicted FRET in the closed MB configuration, we sought to calculate the change in energy transfer if the MB is not correctly formed. A mimic to represent the incorrectly assembled MB would be a denatured MB where the DNA is enzymatically degraded. To that end, the DNase enzyme (a known nuclease) was used to prepare a degraded MB sample (referred to as degraded-MB). To ensure the microenvironment around the dyes remains the same when comparing intact MB with degraded MB, the intact MB was spiked with deactivated DNase enzyme at the same concentration.

Using the vendor-recommended protocol, intact and degraded-MB samples were assembled. Briefly, the degraded-MB sample PCR tube contained \qty{10}{\micro\L} DNase buffer, \qty{1}{\micro\L} DNase (normal), \qty{1}{\micro\L} of MB, and \qty{77.57}{\micro\L} water to bring the reaction volume to \qty{100}{\micro\L}. DNase digestion was achieved by incubating the sample for \qty{10}{\minute} at \qty{37}{\degree C} (as recommended by the vendor) followed by enzyme deactivation for \qty{15}{\minute} at \qty{75}{\degree C} and cooldown to room temperature (\qty{25}{\degree C}). Finally, \qty{1}{\micro\L} of \qty{0.5}{\Molar} EDTA pH 8.0 was added to ensure complete DNase loss of activity. Intact-MB (but in the same enzymatic environment) was prepared by adding all the above-mentioned solutions in a PCR tube except for the MB itself. The MB was added after the incubating as mentioned above, adding \qty{1}{\micro\L} of \qty{0.5}{\Molar} EDTA.

\qty{15}{\micro\L} of both samples were loaded onto a 96-well plate and into a TECAN Spark plate reader. The fluorescence spectra were measured using an excitation wavelength of \qty{420}{\nm} and a bandwidth of \qty{10}{\nm}, from \qtyrange{475}{800}{\nm}. \textbf{Figure~\ref{fig:S2}} shows the steady-state change in fluorescence measured for the intact and degraded MB.

\subsection*{S4.2 Results for the degraded/Intact MB}
\begin{figure}[h!]
    \centering
    \includegraphics[width=\textwidth]{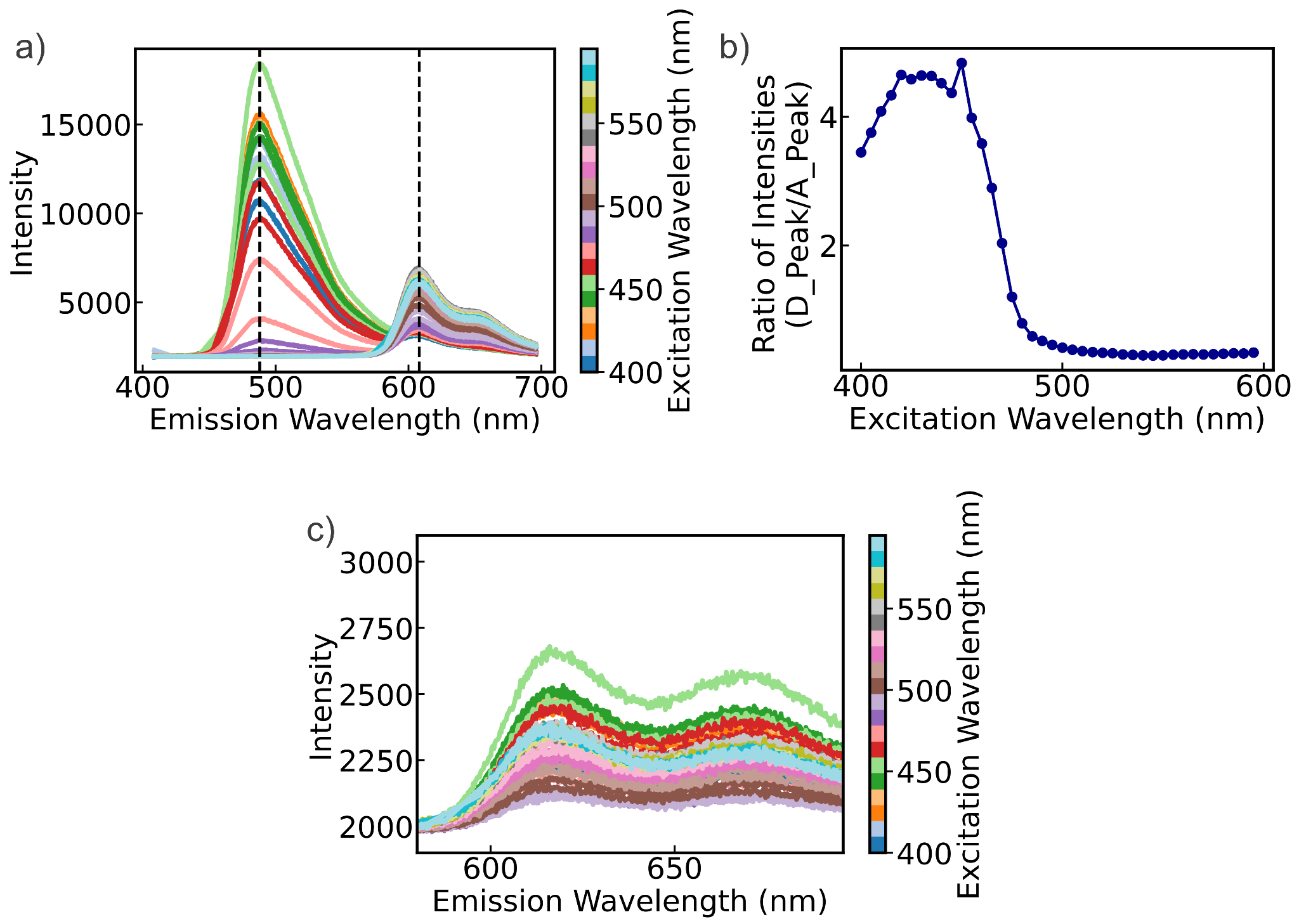} 
    \caption{Beacon denaturing results a)FL emission for the test (degraded) sample b) How the donor emission beak changed with increasing excitation for the test (degraded) sample c) FL emission for the control (intact) sample.}
    \label{fig:S2}
\end{figure}

Fluorescent emission spectroscopy was conducted on the degraded-MB prepared as described in the Sample Preparation section. The resulting emission spectrum is presented in \textbf{Figure~\ref{fig:S2}}

The emission spectrum for the degraded sample shows two main peaks with varying intensity levels when the sample is excited at wavelengths between \qtyrange{400}{595}{\nm} (refer to \textbf{Figure~\ref{fig:S2}a}). The peaks are observed at \qty{488}{\nm} (donor dye, Atto 425) and \qty{610}{\nm} (acceptor dye, Cy3.5). As shown in \textbf{Figure~\ref{fig:S2}b} the emission intensity increases with the excitation wavelength up to a certain point, after which it begins to decrease, consistent with the excitation reaching the peak absorption of the donor and acceptor dyes. The presence of two distinct emission peaks indicates that the beacon denaturing process was successful. The spectral output closely resembles that of two separate dyes in solution, confirming that the donor and acceptor dyes are effectively behaving as independent emitters in the denatured beacon sample.

Fluorescent emission spectroscopy was also performed on the control (intact-MB) beacon sample, and the resulting emission spectrum is presented in \textbf{Figure~\ref{fig:S2}c}. The emission spectrum of the intact MB sample displays two peaks: a prominent peak at \qty{615}{\nm} and a shorter peak at \qty{671}{\nm}. According to the literature, Cy3.5 exhibits a major peak at \qty{591}{\nm} and a secondary peak at \qty{640}{\nm}, which corresponds to an intermediate emission state. The redshift of the peaks observed here is consistent with the literature data. Importantly, no significant peak is observed at \qty{488}{\nm}, which is the expected emission peak for the donor dye, Atto425. This absence suggests that the donor dye is transferring its energy to the acceptor dye, which emits light. These results confirm the occurrence of Förster Resonance Energy Transfer (FRET) between the donor and acceptor dyes. The presence of FRET in the intact beacon sample demonstrates the successful design and functionality of the molecular beacon, indicating that the donor and acceptor are in the desired spatial arrangement to facilitate energy transfer.
\section*{S5. Supplemental Data for Opened-MB Analysis}

\subsection*{S5.1 Stochastic Baseline}
The opened MB was prepared by combining the MB with 5 times the amount of its complementary strand in the same buffer conditions (PBS buffer). The solution was then heated to \qty{90}{\degree C} and naturally cooled in the heating block to room temperature ($\sim\qty{2}{\hour}$). The 5 times excess of complementary strands ensures that the total population of closed MB will likely interact with the complementary oligos, leaving a negligible population of uninteracted MB in the sample.

\begin{figure}[h!]
    \centering
    \includegraphics[width=\textwidth]{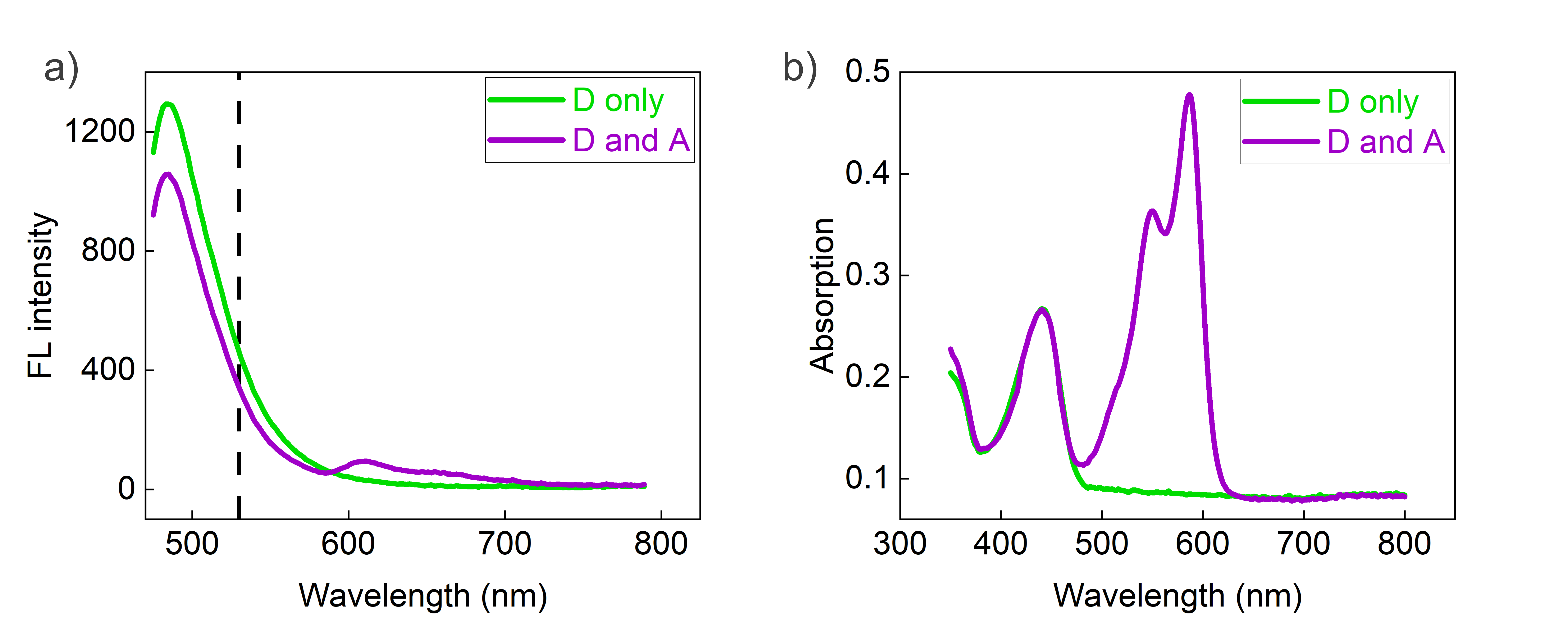} 
    \caption{Steady state FL results for the complimentary beacon analysis (opened-MB)}
    \label{fig:S3}
\end{figure}

\textbf{Figure~\ref{fig:S3}} illustrates the results of FRET following the complementary beacon analysis. As shown in the absorbance spectrum (\textbf{Figure~\ref{fig:S3}b}), the concentrations of the donor-only sample and the donor-acceptor sample are comparable. This similarity allows for a valid comparison of fluorescence emission intensities under steady-state conditions to calculate the FRET efficiency.

Steady-state fluorescence emission spectroscopy is shown in \textbf{Figure~\ref{fig:S3}a}, and the FRET efficiency was estimated based on the donor fluorescence intensity at \qty{532}{\nm}, the region of interest in this study. The calculated FRET efficiency was approximately 21\%. 

\section*{S6. Numerical Fitting and Lifetime analysis}

\subsection*{S6.1 Multiexponential Decay numerical fitting procedure}
Numerical decay curves for both the donor-only and donor-acceptor samples were analyzed using a multi-exponential decay model. The instrument response function (IRF) of the TCSPC system was measured using a dilute colloidal silica scattering solution (Ludox 1:100 dilution), yielding a Full Width at Half Maximum (FWHM) of \qty{5}{\ps}. Since the characteristic decay times ($\tau_i$) of both samples were significantly longer than the IRF pulse width, a tail-fitting approach was employed. In this method, the fitting range was initiated at the peak of the PL intensity to minimize the influence of the excitation pulse and instrumental artifacts on the extracted parameters.

The experimental data were fitted using the following multiexponential function:
\begin{equation}
y(t) = y_0 + \sum_{i=1}^{n} A_i \exp\left(-\frac{t - t_0}{\tau_i}\right)
\end{equation}
Where $y_0$ is the baseline/dark count, $A_i$ is the pre-exponential (amplitude) of the $i$-th decay component, $\tau_i$ is the associated decay lifetime, and $t_0$ is the time shift constant.

The optimal number of exponential terms ($n$) was determined by evaluating the reduced chi-squared ($\chi^2$) values. A fit was considered statistically robust when $\chi^2$ approached unity ($0.9 < \chi^2 < 1.3$) and no signal was observed in the residuals plot beyond a random noise distribution.

\begin{table}[h!]
\centering
\caption{Associated decay times, amplitudes and chi-squared for the closed-MB with donor and acceptor}
\label{tab:S2}
\begin{tabular}{lccccccc}
\toprule
Closed-MB & $\tau_1$ (\unit{\ns}) & $A_1$ & $\tau_2$ (\unit{\ns}) & $A_2$ & $\tau_3$ (\unit{\ns}) & $A_3$ & $\chi^2$ \\
\midrule
On glass & $0.231 \pm 0.013$ & $600.89$ & $2.018 \pm 0.116$ & $3919.93$ & $6.219 \pm 1.257$ & $1939.55$ & $0.999$ \\
On ENZ & $0.054 \pm 0.001$ & $2410.93$ & $0.532 \pm 0.012$ & $1705.48$ & $2.834 \pm 0.016$ & $3544.69$ & $0.999$ \\
\bottomrule
\end{tabular}
\end{table}

\begin{table}[h!]
\centering
\caption{Associated decay times, amplitudes and chi-squared for the donor-MB (opened configuration)}
\label{tab:S3}
\begin{tabular}{lccccc}
\toprule
Opened donor-MB & $\tau_1$ (\unit{\ns}) & $A_1$ & $\tau_2$ (\unit{\ns}) & $A_2$ & $\chi^2$ \\
\midrule
On glass & $0.778 \pm 0.019$ & $1425.41$ & $3.109 \pm 0.210$ & $5153.07$ & $0.999$ \\
On ENZ & $0.398 \pm 0.013$ & $2875.03$ & $2.882 \pm 0.055$ & $3825.34$ & $0.998$ \\
\bottomrule
\end{tabular}
\end{table}

\begin{table}[h!]
\centering
\caption{Associated decay times, amplitudes and chi-squared for the opened-MB with donor and acceptor}
\label{tab:S4_decay}
\begin{tabular}{lccccccc}
\toprule
Opened-MB (donor+acceptor) & $\tau_1$ (\unit{\ns}) & $A_1$ & $\tau_2$ (\unit{\ns}) & $A_2$ & $\tau_3$ (\unit{\ns}) & $A_3$ & $\chi^2$ \\
\midrule
On glass & $0.337 \pm 0.010$ & $2058.14$ & $1.623 \pm 0.101$ & $2782.45$ & $4.860 \pm 0.502$ & $2131.53$ & $0.999$ \\
On ENZ & $0.082 \pm 0.003$ & $2107.46$ & $0.553 \pm 0.010$ & $2740.44$ & $3.175 \pm 0.024$ & $2951.82$ & $0.999$ \\
\bottomrule
\end{tabular}
\end{table}

\subsection*{S6.2 Amplitude Averaged Decay Behavior}
Calculations for the amplitude weighted average lifetime and average decay rates ($\Gamma$) for the entire system are given by the following equations:
\begin{equation}
\langle \tau \rangle = \frac{\sum_{i} A_i \tau_i}{\sum_{i} A_i}
\end{equation}
\begin{equation}
\langle \Gamma \rangle = \frac{1}{\langle \tau \rangle}
\end{equation}

\begin{table}[h!]
\centering
\caption{Associated amplitude-weighted average decay times and decay rates for each system for ensemble analysis}
\label{tab:S4_avg}
\begin{tabular}{lcc}
\toprule
System & \begin{tabular}[c]{@{}c@{}}$\langle \tau \rangle$ (\unit{\ns})\\ Amplitude-weighted avg.\end{tabular} & \begin{tabular}[c]{@{}c@{}}$\langle \Gamma \rangle$ (\unit{\ns^{-1}})\\ Decay rate\end{tabular} \\
\midrule
Closed-MB on glass & 3.113 & 0.321 \\
Closed-MB on ENZ & 1.447 & 0.691 \\
Opened donor-MB on glass & 2.604 & 0.384 \\
Opened donor-MB on ENZ & 1.816 & 0.551 \\
Opened-MB on glass & 2.233 & 0.448 \\
Opened-MB on ENZ & 1.418 & 0.705 \\
\bottomrule
\end{tabular}
\end{table}

The amplitude weighted average decay times are presented here for completeness. The component-specific analysis in the main text is the primary interpretation for the fluorescence decay behavior due to the qualitative change of an additional fast decay channel in the ENZ-modified system specific to the interacting donor fraction. Alternatively, the amplitude-weighted average lifetimes presented in \textbf{Table~\ref{tab:S4_avg}} provide a holistic representation of the total molecular ensemble behavior, albeit while over-representing the slow decay behavior. Although this analysis likely includes the behavior of low- or non-interacting populations, a consistent photonic enhancement is nevertheless observed across both configurations. Specifically, for the fully functionalized ‘opened-MB’ system, the ensemble-averaged decay rate increases from \qty{0.448}{\ns^{-1}} on glass to \qty{0.705}{\ns^{-1}} on the ENZ substrate. This global acceleration confirms that the ENZ-mediated enhancement significantly affects the global decay behavior of the averaged emitter population.

\subsection*{S6.3 Confirmation of FRET-specific enhancement}
For a donor-acceptor emitter pair near a metallic or ENZ substrate, or more generally near any photonic system that modifies the local electromagnetic density of states, the donor excited-state lifetime can be altered even in the absence of an acceptor. In our system this can potentially be caused by multiple mechanisms such as increased radiative LDOS, increased non-radiative loss, coupling to lossy modes, or some combination of these effects. Changes in the donor lifetime on the ENZ substrate can modify the baseline decay rate and thus influence the apparent lifetime-based FRET efficiency. In the system presented in the manuscript, an increase in the fluorescent decay rate is observed in the donor-only sample on the ENZ relative to glass substrate. The lifetime-derived FRET rate enhancement should therefore be interpreted separately from the total FRET efficiency baseline to decompose the FRET-specific (or acceptor-modified) effect on the transfer rate.

Note that the measurement is on the donor decay rate. Consider a simple decomposition of the total decay rate into donor-specific ($k_{\text{rad}}$ and $k_{\text{nrad}}$) and FRET-specific ($k_{\text{FRET}}$) components. Since we interpret each decay channel as an independent first-order pathway depleting the same population of donor-excited emitters, these components are additive:
\begin{equation}
k_{\text{DA}} = k_{\text{rad}} + k_{\text{nrad}} + k_{\text{FRET}} = k_{\text{D}} + k_{\text{FRET}}
\end{equation}

Thus, the contribution of the FRET-based component of the total decay is given by
\begin{equation}
k_{\text{FRET}} = k_{\text{DA}} - k_{\text{D}}
\end{equation}
for the donor-acceptor $k_{\text{DA}}$ and donor-only $k_{\text{D}}$ measurements. The enhancement factor can be stated as
\begin{equation}
\text{Enhancement Factor} = \frac{k_{\text{FRET, ENZ}}}{k_{\text{FRET, Glass}}} = \frac{k_{\text{DA, ENZ}} - k_{\text{D, ENZ}}}{k_{\text{DA, Glass}} - k_{\text{D, Glass}}}
\end{equation}

\paragraph{For the closed-MB}
By looking at the decay channels in each case stated in \textbf{Table~\ref{tab:S2}}. \\
\textbf{Assumptions:} $k_{\text{D,Glass}}$ and $k_{\text{D,ENZ}}$ for the donor-only system are considered the same as in the opened-MB case. This is reasonable based on the emitter geometries. \\
By looking at the individual channels ($\tau_1$): \\
Enhancement factor ($\tau_{1\text{\_closed}}$) = 5.25

\paragraph{For the opened-MB}
By looking at the decay channels in each case stated in \textbf{Table~\ref{tab:S3}} and \textbf{Table~\ref{tab:S4_decay}}, \\
By looking at the individual channels ($\tau_1$): \\
Enhancement factor ($\tau_{1\text{\_opened}}$) = 5.67

\begin{figure}[h!]
    \centering
    \includegraphics[width=\textwidth]{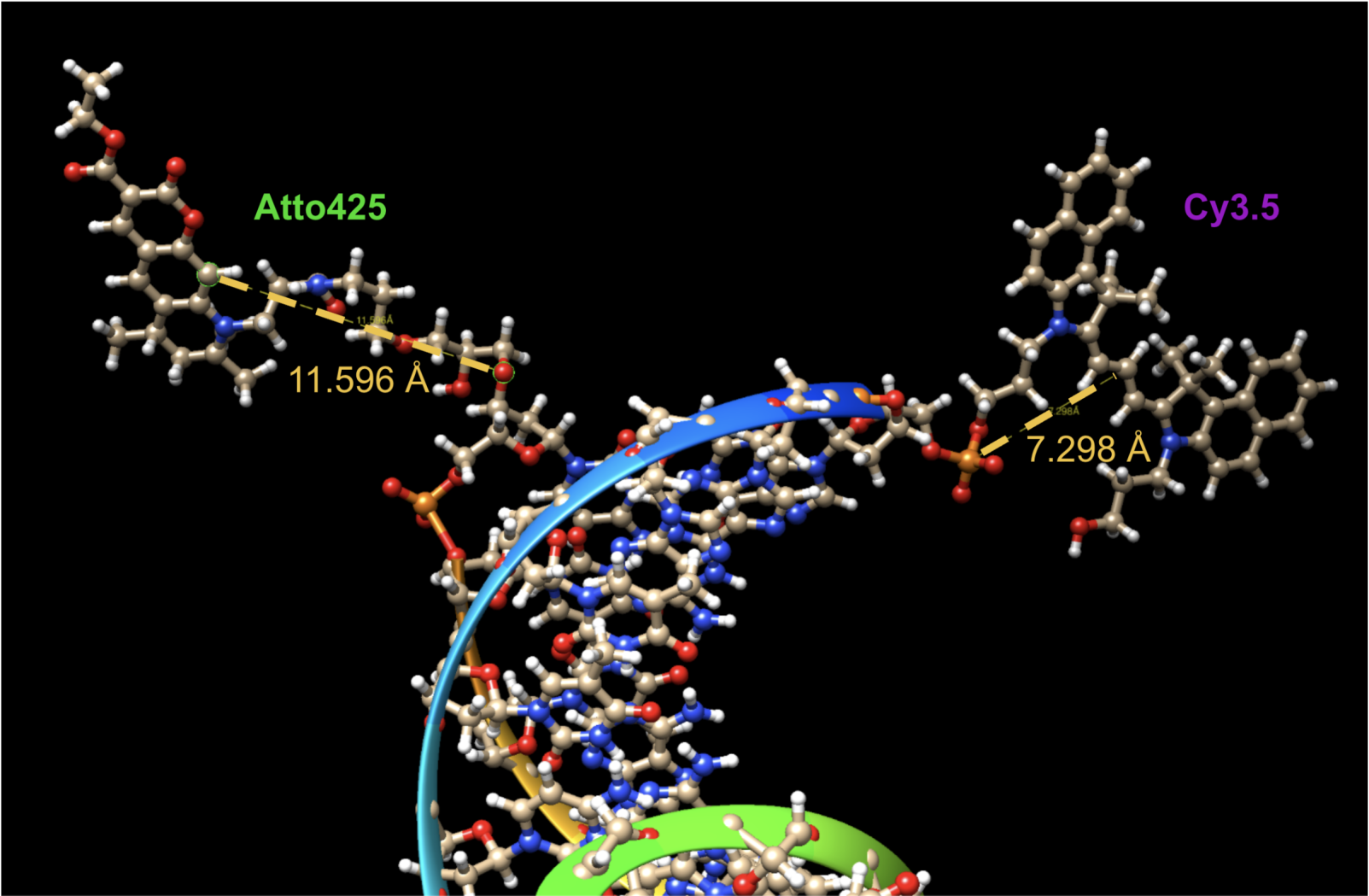} 
    \caption{Structural modeling of the closed-MB: The PDB structure was visualized and analyzed using UCSF Chimera and AlphaFold3 to calculate the effective tethered length of the 3' end to Atto425 and 5' end to Cy3.5 fluorophore attachments}
    \label{fig:S4}
\end{figure} 

\section*{S7. Theoretical FRET parameters}

\subsection*{S7.1 Predicted Donor-Acceptor Distances}
The DNA-programmed architectures provide structurally enforced separations ($r$) between the Atto 425 and Cy3.5 dye molecules.

\textbf{Closed-MB:} The hairpin stem loop structure brings the dyes into a predicted proximity of $\sim\qty{2}{\nm}$

\textbf{Opened-MB:} Upon hybridization with the complementary strand, the rigid 24-base duplex results in a predicted separation of \qty{8.16}{\nm} ($\qty{0.34}{\nm} \times 24$), corresponding to the contour length of the DNA helix.

The estimated distances from the ends of the MB to the fluorophores are stated below and illustrated in \textbf{Figure~\ref{fig:S4}}.

The estimated distances: \\
3' end to Atto425 $\sim\qty{11.596}{\angstrom}$ \\
5' end to Cy3.5 $\sim\qty{7.298}{\angstrom}$

\subsection*{S7.2 Förster Radius ($R_0$)}
The theoretical efficiency was benchmarked against standard solution parameters.

\textbf{Förster Radius ($R_0$):} The $R_0$ for the Atto 425/Cy3.5 pair is calculated to be $\sim\qty{5.77}{\nm}$ in aqueous solution.

\textbf{Stochastic efficiency:} Using the equation,
\begin{equation}
E = \frac{1}{1 + \left(\dfrac{r}{R_0}\right)^6}
\end{equation}

The predicted efficiency for the \qty{8.16}{\nm} opened-MB in the absence of photonic enhancement is $\sim 21\%$.

\subsection{S7.3 Linker Conformational Freedom and Donor-Acceptor Separation Distributions}

\begin{figure}[h!]
    \centering
    \includegraphics[width=\textwidth]{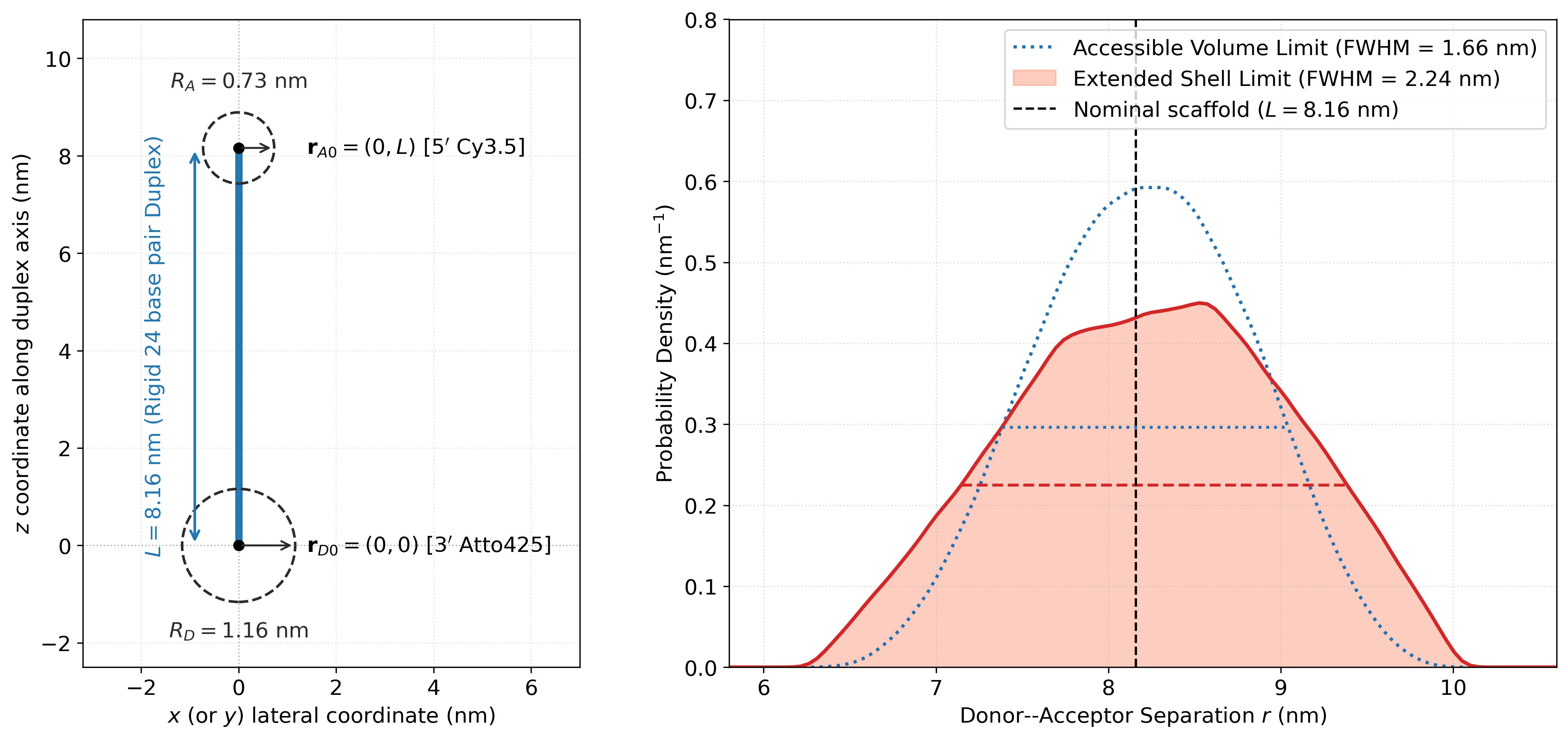}
    \caption{Conformational distance distributions from terminal linkers. 
    Left: Two-dimensional Cartesian representation of the opened duplex ($L = 8.16$~nm, blue rod) with conformational bounds defined by the donor ($R_D = 1.16$~nm) and acceptor ($R_A = 0.73$~nm) tethering linkers. 
    Right:Corresponding donor-acceptor distance distributions comparing the accessible volume limit (blue dotted curve, $\text{FWHM} = 1.66$~nm; $\langle r \rangle_{\mathrm{FRET}} = 8.08$~nm) and the extended shell limit (shaded red area, $\text{FWHM} = 2.24$~nm; $\langle r \rangle_{\mathrm{FRET}} = 8.00$~nm). Black dashed line indicates the nominal $8.16$~nm scaffold length; horizontal lines mark the respective full-widths at half-maximum.}
    \label{fig:S4_b}
\end{figure}

While the rigid 24-bp duplex defines a nominal contour separation of $L = 8.16$~nm (well within the $\sim 50$~nm double stranded DNA persistence length),\supercite{sindbert2011accurate} the aliphatic linkers tethering the $3^\prime$-Atto425 ($R_D = 1.16$~nm) and $5^\prime$-Cy3.5 ($R_A = 0.73$~nm) introduce conformational variations (\textbf{Figure~\ref{fig:S4_b}}, left). To quantify this effect, we evaluated the spatial separation distribution across two physical extremes (\textbf{Figure~\ref{fig:S4_b}}, right):

\begin{enumerate}
    \item \textbf{Accessible Volume Limit (Solid Sphere):} Uniform sampling over the accessible tethered volume yields an arithmetic mean distance of $\langle r \rangle = 8.21 \pm 0.61$~nm ($\text{FWHM} = 1.66$~nm). Due to the steep $r^{-6}$ distance dependence of dipole-dipole coupling, the effective FRET-weighted distance is $\langle r \rangle_{\mathrm{FRET}} = (\langle r^{-6} \rangle)^{-1/6} \approx 8.08$~nm.
    \item \textbf{Extended Shell Limit (Spherical Surface):} Constraining the dyes to fully extended swiveling rigid linkers broadens the distribution ($\text{FWHM} = 2.24$~nm, spanning $6.27\text{-}10.05$~nm), shifting the effective distance to $\langle r \rangle_{\mathrm{FRET}} \approx 8.00$~nm.
\end{enumerate}

In both bounding limits, the effective FRET-weighted separation deviates from the nominal $8.16$~nm value by $\le 2\%$. Because our numerical Green's tensor simulations indicate that ENZ-induced transfer acceleration remains robust between $4\times$ and $8\times$ across the entire $5\text{-}10$~nm separation window (\textbf{Figure~5c} and \textbf{Figure~S9}), the nominal scaffold length provides an accurate microscopic representation of the ensemble transfer dynamics.

\section*{S8. Simulation of substrate-mediated FRET enhancement}

\begin{figure}[h!]
    \centering
    \includegraphics[width=0.5\textwidth]{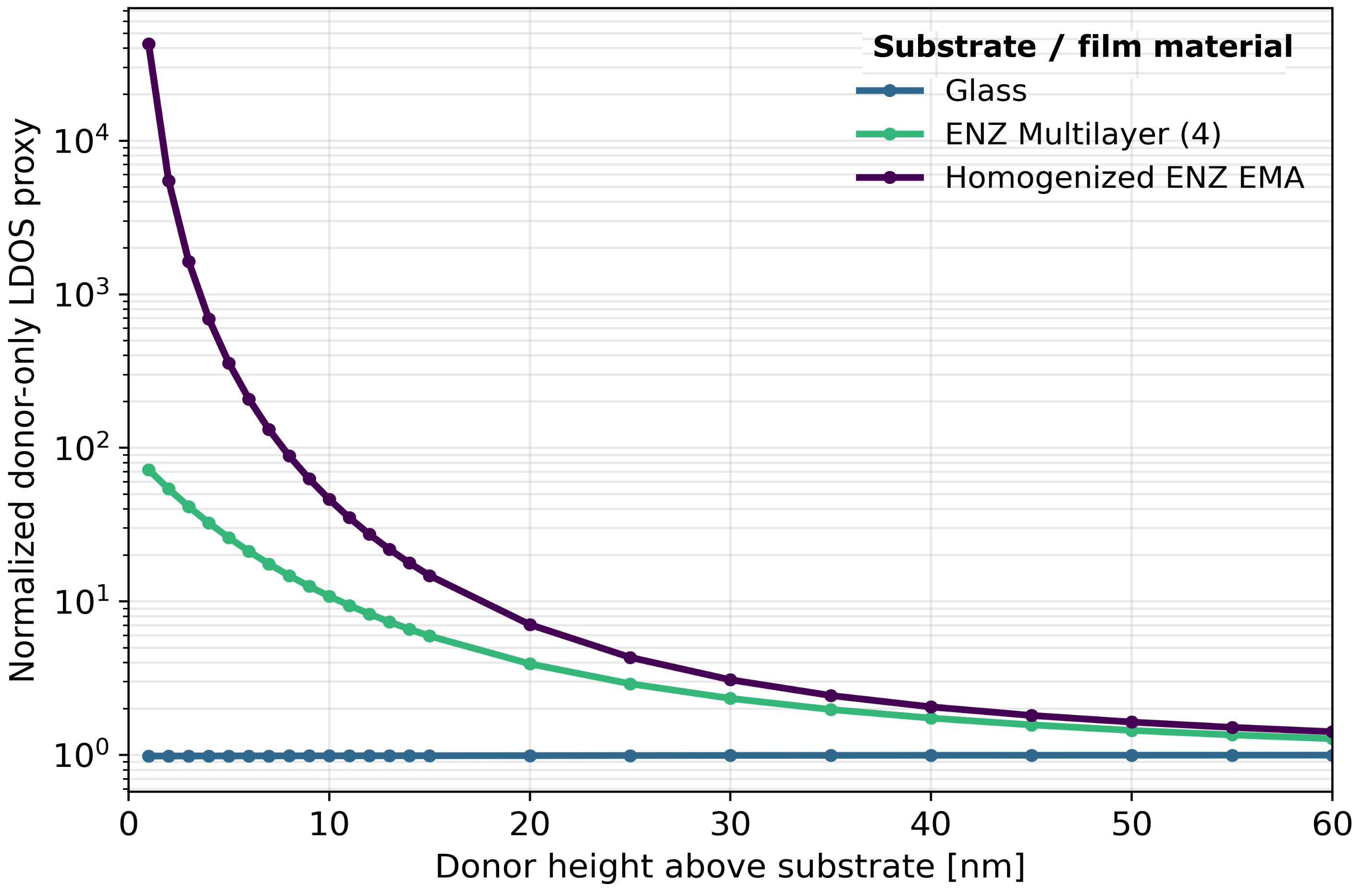} 
    \caption{Maximum normalized donor-only orientation-resolved LDOS (decay rate) proxy versus donor height above the substrate for glass, explicit ENZ multilayer, and homogenized ENZ effective-medium substrates. The ENZ structures strongly enhance the donor-only orientation-resolved LDOS proxy near the interface, while glass produces minimal modification.}
    \label{fig:S5}
\end{figure}

\subsection*{S8.1 Model and Problem Geometry}
To interpret the experimental observation of ENZ substrate-mediated donor-acceptor coupling, electromagnetic Green tensor simulations are performed for a donor-acceptor dye pair embedded in a PVA superstrate half-space above a planar substrate \supercite{Dung2002,Zangwill2013,sipe1987new,Tomas1995}. The dye molecules are modeled as point dipoles centered on the transition dipole moment vector, and the effect of the DNA nanostructure is only implicit in the dipole spacing and orientation conditions. The optical properties of the DNA are not modeled explicitly since negligible optical scattering is expected. The model geometry schematic is shown in \textbf{Figure~\ref{fig:fig5}a}. The planar substrate interface is defined at $z = \qty{0}{\nm}$, with the PVA layer occupying $z > \qty{0}{\nm}$ and the multilayer or homogenized ENZ layer and glass substrate halfspace occupying $z < \qty{0}{\nm}$. The PVA film is treated as a semi-infinite superstrate. This approximation neglects the upper PVA / air interface; however, the effect on the near-substrate molecular population was observed to be insignificant. The donor and acceptor are treated as classical point dipoles with definite positions $\mathbf{D}$ and $\mathbf{A}$, respectively, within the PVA media. The donor-acceptor separation is denoted by the vector $\mathbf{s}$, which is parameterized by a midpoint height $h$ above the substrate and the pair-axis orientation $\theta_{\text{pa}}$, which is the polar angle of the separation vector with the surface normal. The donor and acceptor transition dipole orientations are described by $\theta_{\text{dd}}$ for the donor dipole polar angle with respect to the surface normal, $\theta_{\text{da}}$ for the angle between the donor and acceptor transition dipoles, and $\theta_{\text{az}}$ and $\theta_{\text{dz}}$ for the azimuthal angle of acceptor and donor, respectively, at fixed $\theta_{\text{da}}$ and $\theta_{\text{dd}}$. Due to the rotational symmetry of the planar substrate about the surface normal, the azimuthal orientation of the separation vector can be disregarded without loss of generality.

The electromagnetic Green tensor between the donor and acceptor is given as
\begin{equation}
\mathbf{G}(\mathbf{r}_{\mathbf{A}}, \mathbf{r}_{\mathbf{D}}, \omega) = \mathbf{G}_0(\mathbf{r}_{\mathbf{A}}, \mathbf{r}_{\mathbf{D}}, \omega) + \mathbf{G}_{\text{ref}}(\mathbf{r}_{\mathbf{A}}, \mathbf{r}_{\mathbf{D}}, \omega)
\end{equation}
where $\mathbf{G}_0$ is the homogeneous Green tensor in the PVA matrix and $\mathbf{G}_{\text{ref}}$ is the scattered or reflected Green tensor from the substrate. Since the substrate consists of a single or multilayer planar stack, the reflected Green tensor can be calculated via the reflected Fresnel coefficients. This analysis includes both the propagating and evanescent wavevectors. Note that this formulation employs coherent interference between the direct donor field and the substrate-reflected fields; the substrate can therefore either enhance or suppress the donor-acceptor transfer metric depending on height, separation, and dipole orientation. Note that this analysis is conducted at 532~nm near the donor - acceptor overlap peak, and neglects the remainder of the overlap band including variance of the Green tensor in this range, which can contribute to the experimental observations. In the donor-only simulation, the relevant metric is the enhancement of the orientation-resolved LDOS proxy normalized a homogeneous system, which is given as
\begin{equation}
\text{LDOS}_{\text{proxy}} = 1 + \frac{6\pi}{k} \operatorname{Im}\left[\boldsymbol{\mu} \cdot \mathbf{G}_{\text{ref}}(\mathbf{r}_{\mathbf{D}}, \mathbf{r}_{\mathbf{D}}, \omega) \cdot \boldsymbol{\mu}\right]
\end{equation}
where $\boldsymbol{\mu}$ is the donor transition dipole, and $k$ is the wave number in the emitter host media. The homogeneous component evaluates to unity in the ratio, and the second term is the substrate-induced oriented LDOS correction via the reflected field. Since the donor and acceptor molecules are located in spatially distinct positions during energy transfer processes, the unnormalized electromagnetic transfer amplitude must consider the entire Green tensor, and is given as
\begin{equation}
A_{\text{transfer}} = \boldsymbol{\mu}_{\mathbf{A}} \cdot \mathbf{G}(\mathbf{r}_{\mathbf{A}}, \mathbf{r}_{\mathbf{D}}, \omega) \cdot \boldsymbol{\mu}_{\mathbf{D}}
\end{equation}
where $\boldsymbol{\mu}_{\mathbf{D}}$ and $\boldsymbol{\mu}_{\mathbf{A}}$ are the transition dipoles for the donor and acceptor, respectively. The relevant transfer metric is the intensity of the transfer field
\begin{equation}
M = |A_{\text{transfer}}|^2.
\end{equation}
The relevant normalized coupling enhancement is then calculated as $\eta = M / M_0$, where $M_0$ is the corresponding electromagnetic transfer metric, defined above, for a donor-acceptor dye pair at fixed orientation in the homogeneous medium far (\qty{50}{\nm}) from the surface-interacting volume (and thus without a significant substrate contribution). This normalization definition is utilized so configurations with intrinsically small transfer coupling do not overrepresent in the enhancement metric. However, it should be noted that this metric reports enhancement with respect to a constant homogeneous configuration and not a direct enhancement of the specified configuration.

Note that these simulations report the electromagnetic contribution of the planar photonic environment to the donor-acceptor coupling properties and do not attempt to compute an absolute molecular FRET rate. What is reported are the relative changes in the electromagnetic transfer coupling compared to a constant homogeneous baseline. This treatment captures the substrate-mediated radiative and near-field electromagnetic coupling but does not include other molecular-scale effects such as dye-specific conformational disorder and heterogeneous emitter populations, which are clearly present in the experimental system. Additionally, unless otherwise stated, all curves report the maximum value of the coupling metric over all remaining free parameters.

\begin{figure}[h!]
    \centering
    \includegraphics[width=\textwidth]{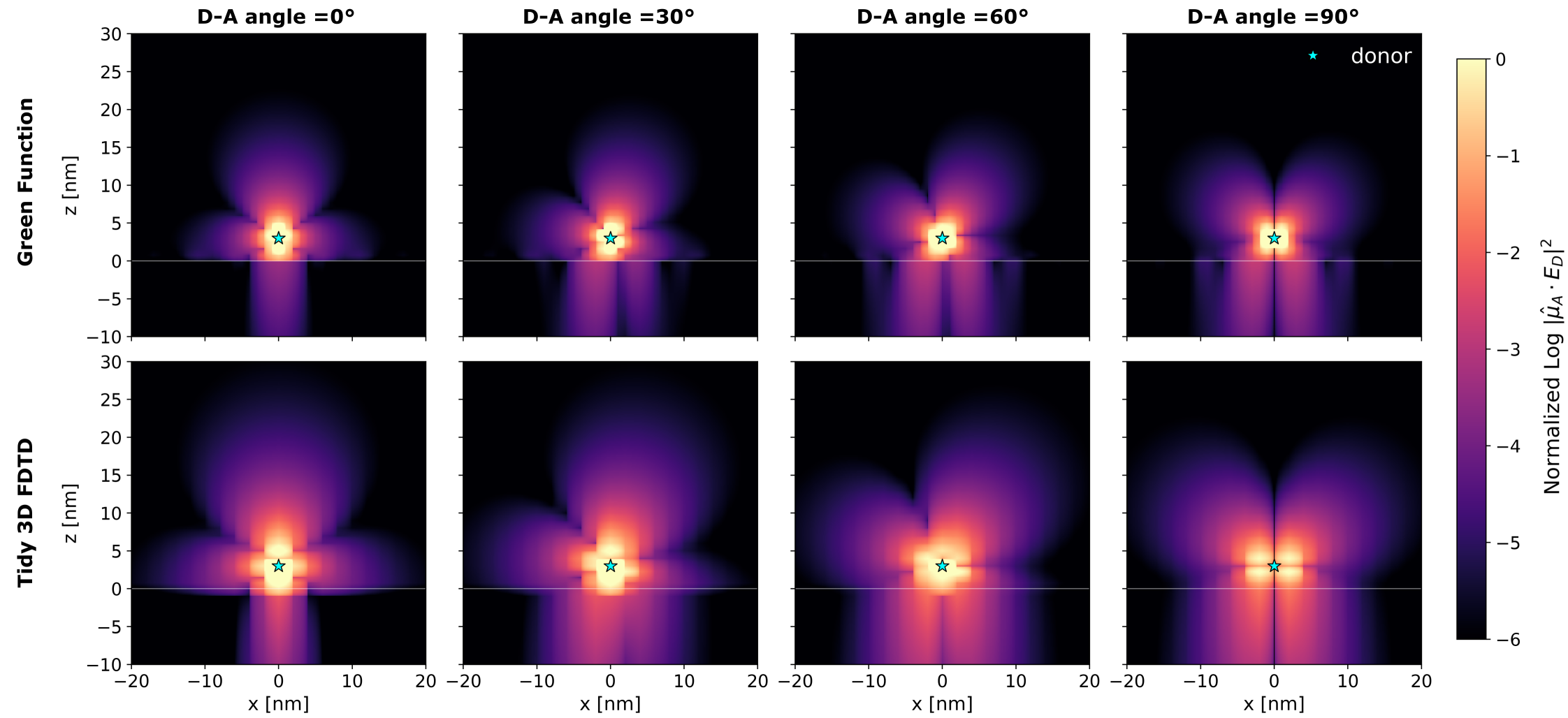} 
    \caption{Spatial maps of the normalized logarithm of the transfer field between a donor dipole at $z = \qty{3}{\nm}$ (blue star) and an acceptor dipole at various spatial positions. The column headers denote the donor-acceptor angle considered for the test acceptor dipole. The top row shows results from the Green tensor formalism described above, and the bottom row shows results for the FDTD simulations performed in Tidy3D software. The ENZ substrate is located in $z \le \qty{0}{\nm}$. This comparison demonstrates both the similarity between the Green tensor and Tidy3D results and the dependence on donor-acceptor pair orientation for enhancement when considering specific dipole orientations. All plots show the maximum of the transfer field with respect to all remaining free degrees of orientational freedom.}
    \label{fig:S6}
\end{figure}

\subsection*{S8.2 Donor-only decay modification}
\textbf{Figure~\ref{fig:S5}} shows the normalized donor-only orientation-resolved LDOS proxy as a function of donor height above the substrate. This simulation only takes into account a single dipole rather than the FRET pair. Three substrates are considered: homogeneous glass; an explicit multilayer ENZ film stack containing the four \ce{Au}/\ce{TiO2} bilayer structure discussed in the manuscript; and a homogenized ENZ film matching the total stack thickness with the permittivity matched to an anisotropic effective medium, also as shown in the manuscript. The glass reference remains near unity when approaching the substrate, indicating a minimal modification to the donor decay rate relative to homogeneous PVA. In contrast, both the multilayer and homogeneous ENZ films (stacks) produce a strong near-surface increase in the normalized donor-only orientation-resolved LDOS proxy. The enhancement is largest within the first few nanometers from the surface; however, significant enhancement is still observed on the order of tens of nanometers from the interface.

The ENZ substrate is expected to modify donor-only decay even in the absence of the acceptor, and furthermore, the magnitude of the modification is felt throughout the entire film, but more acutely at the surface. Note that the penetration length of the donor-only effect into the film occurs over a longer distance than the transfer enhancement effect, which will be discussed below.

\begin{figure}[h!]
    \centering
    \includegraphics[width=\textwidth]{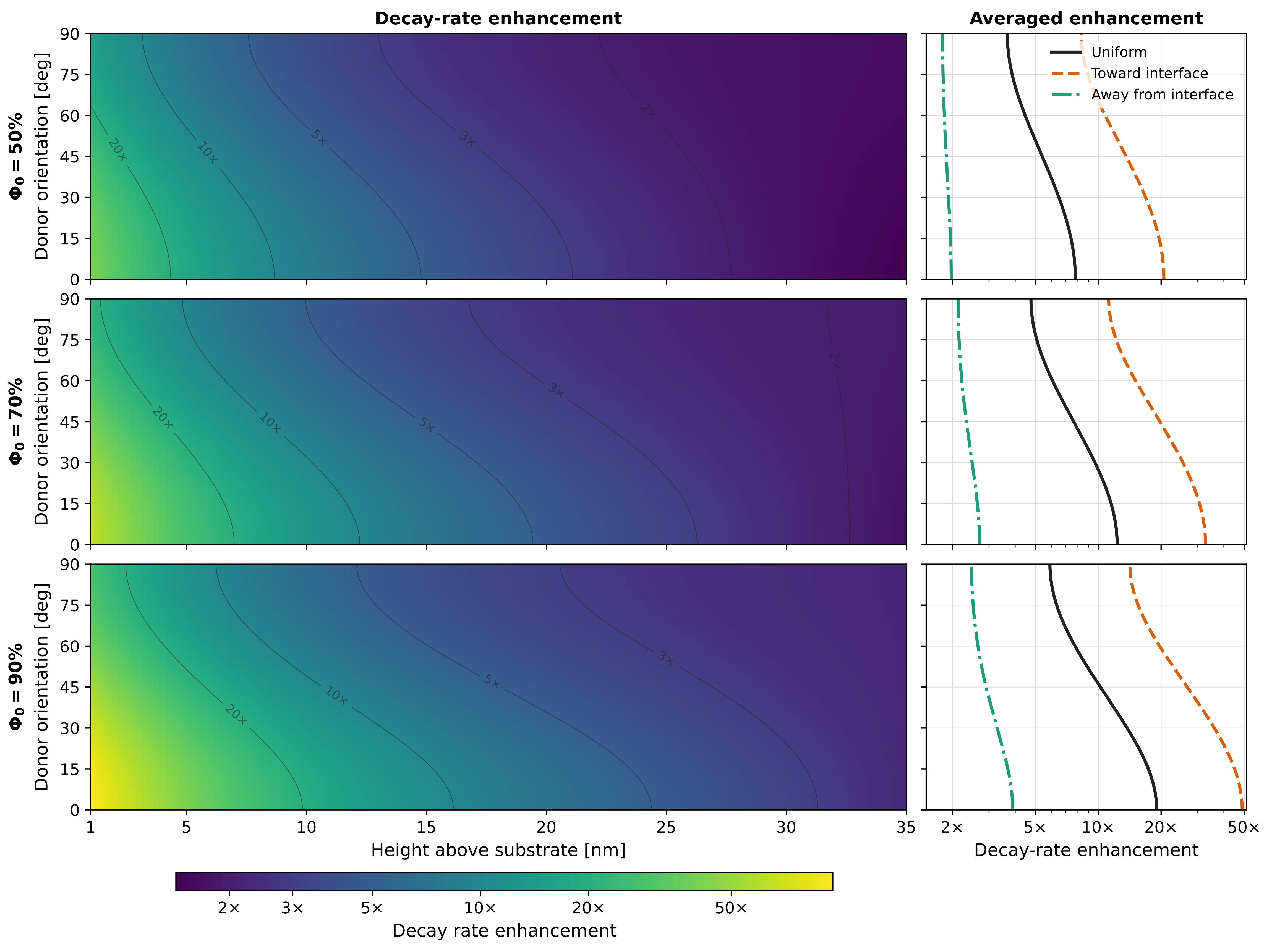} 
    \caption{Comparison of the decay rate enhancement on glass and ENZ substrates for a single dipole (donor-only configuration). Each of the three rows utilize a different value for the donor quantum yield from 50 - 90\%. The left column plots the decay rate enhancement as a function of donor dipole orientation with respect to the substrate interface and the height above the substrate interface into the PVA film. Substantial enhancement factors are observed for dipoles in close proximity to the substrate. The right column shows the height-averaged decay rate enhancement profiles for each considered quantum yield, using three spatial averaging techniques: a uniform distribution of emitters; an exponential distribution weighted toward the substrate interface with $\tau=5$~nm; and a similar distribution with the same extent weighted toward the top of the PVA film away from the substrate.}
    \label{fig:S7}
\end{figure}

\subsection*{S8.3 Angular dependence of substrate-mediated transfer enhancement}
The angular dependence of the normalized coupling enhancement is shown in \textbf{Figure~\ref{fig:fig5}b} for the entire range of donor dipole and pair-axis orientations, respectively, at a fixed position and separation (\qty{8}{\nm} dipole spacing and a midpoint fixed at \qty{3}{\nm} above the surface). In each case, the maximum enhancement as a function of angle is shown as the extreme bounding line, and the shaded region represents the range of enhancements possible by considering the entire range of remaining free spatial and alignment orientation parameters. The ENZ considered in this analysis is the multilayer anisotropic ENZ. The ENZ structure produces a strongly anisotropic optical response, with enhancement depending on both the donor dipole and donor-acceptor pair axis orientations. This is consistent with the anisotropic electromagnetic boundary conditions arising from an ENZ medium. In contrast, the glass substrate shows a far weaker angular dependence.

These results indicate that even when considering a fixed emitter separation and relative dipole orientation angle - the ideal case of a perfectly rigid DNA nanostructure - the orientation of the nanostructure with respect to the surface normal can have a significant impact on the enhancement properties. Since the DNA nanostructure position and orientation are not fixed inside the modeled random PVA matrix, the emitter pair can take all possible positions and orientations, leading to multiple sub-populations with different degrees of substrate-induced enhancement. However, in an experimental system, ordering or non-uniformity of the DNA nanostructures in the PVA matrix could bias the system toward specific enhancement regimes.

\subsection*{S8.4 Height dependence of substrate-mediated transfer enhancement}
Normalized transfer enhancement as a function of donor-acceptor midpoint height for dipole spacings from \qtyrange{6}{10}{\nm} is shown in \textbf{Figure~\ref{fig:fig5}d}. For the ENZ substrate, a strong near-field enhancement is observed when the donor-acceptor pair midpoint is positioned within the first several nanometers from the substrate. The enhancement decreases rapidly with increasing midpoint height and approaches the homogeneous-PVA limit at distances greater than $\sim\qty{10}{\nm}$. This behavior is expected, as the strong substrate-mediated coupling enhancement can be dominated by high-momentum evanescent components, which decay rapidly from the interface. The normalized coupling enhancement is also highly sensitive to the donor-acceptor spacing: for larger emitter spacings, the maximum off-surface enhancement shifts to larger midpoint heights. This is likely a reflection of the geometric constraint on pair orientation and separation, since both emitters must be physically contained in the PVA layer and, when too close to the interface, cannot occupy preferential orientations. For the experimental opened-MB configuration, these results indicate that the substrate-mediated enhancement occurs primarily when the emitter pair occupies the first few nanometers above the ENZ interface. As evidenced by the inset, the magnitude of the enhancement is still highly sensitive to dipole and pair axis orientation, as discussed above. Coupling enhancement in the glass substrate system is largely negligible, with only a modest enhancement observed very close to the interface ($<\qty{2}{\nm}$).

These results support the interpretation that significant coupling enhancement, and thus enhanced transfer efficiency, is possible when operating in the ENZ near-field. Additionally, since the experimental PVA films are tens of nanometers ($\sim\qty{50}{\nm}$) thick, only the fraction of molecular beacons located close to the ENZ interface is expected to experience the strongest transfer coupling enhancement. The simulations therefore support a heterogeneous ensemble picture, rather than a uniform enhancement of all donor-acceptor pairs in the film, which is consistent with the experimental results.

\begin{figure}[h!]
    \centering
    \includegraphics[width=\textwidth]{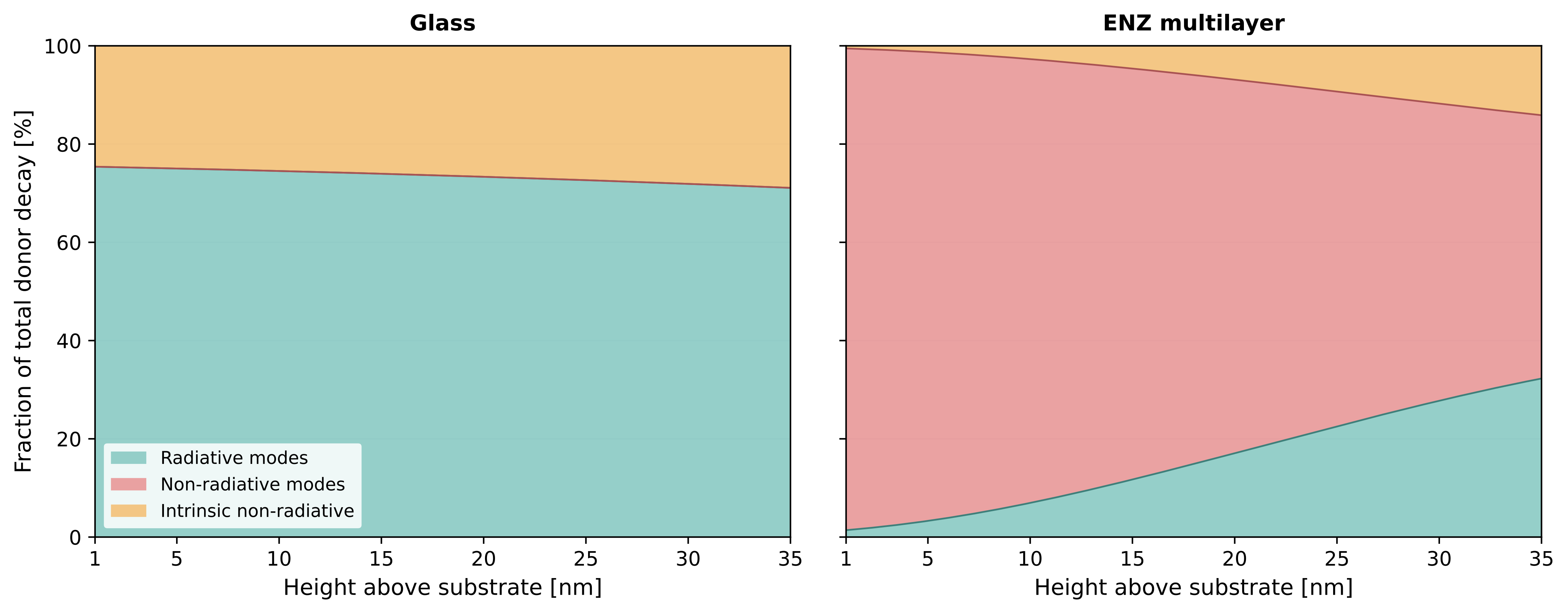} 
    \caption{Analysis of the radiative and non-radiative decay components for a theoretical dipole emitter above glass (left) and ENZ (right) substrates as a function of height above the substrate interface. The donor decay is decomposed into the total radiative modal fraction due to the enhancement from the substrate (using the proxy $\mathrm{k}_{\parallel} \leq \mathrm{k}_{PVA}$), the non-radiative mode fraction, and the intrinsic non-radiative components, using a standard configuration of 45$^{\circ}$ dipole orientation and $\Phi_0=80$\%. The ENZ substrate engenders substantial non-radiative modal fraction.}
    \label{fig:S8}
\end{figure}

\subsection*{S8.5 Validation of the spatial extent}

To confirm the results of the Green tensor simulations, the spatial extent of the transfer field was visualized for a typical donor dipole configuration. The transfer field is defined as the intensity of the E-field component from the donor emission along the acceptor transition dipole for a range of acceptor dipole spatial positions. This visualization was performed for a homogenized \qty{80}{\nm} thick ENZ EMA substrate system. The donor dipole is placed \qty{3}{\nm} from the ENZ interface at $z = \qty{0}{\nm}$, oriented parallel to the surface normal. The normalized transfer field maps, shown in \textbf{Figure~\ref{fig:S6}}, are plotted along the $x$-$z$ interface centered on the donor dipole. First, this visualization utilized the full Green tensor formulation, including the contribution of the nearby ENZ medium through the reflected component. In parallel, the same configuration was generated from a finite-difference time-domain (FDTD) study using Tidy 3D software \supercite{Tidy3D2026}. This simulation utilized identical conditions and geometry, with PML boundary conditions. For the anisotropic ENZ EMA substrate, the negative real components were represented with a one-pole Drude fit to improve the FDTD stability. A \qty{1}{\nm} dimension meshing was utilized near the donor dipole position to better resolve the rapidly changing field structure near the dipole. Finally, for both the Green tensor and FDTD simulations, the transfer field was calculated and shown for a range of acceptor dipole orientations from collinear to orthogonal.

In \textbf{Figure~\ref{fig:S6}}, a close correspondence between the transfer field maps from the Green tensor and FDTD simulations, respectively, is observed. This includes the asymmetry in the $+x$ and $-x$ directions, which is a consequence of the acceptor dipole orientation in the $+x$ direction. In FDTD simulations, a small loss of spatial resolution and associated spatial smearing of the intensity are expected due to the finite meshing and rapidly varying field at short distances. These transfer field maps both provide a validation of the Green tensor formalism and results as well as demonstrate the large effect of spatial separation on transfer enhancement, even very locally to the ENZ interface.

\begin{figure}[h!]
    \centering
    \includegraphics[width=\textwidth]{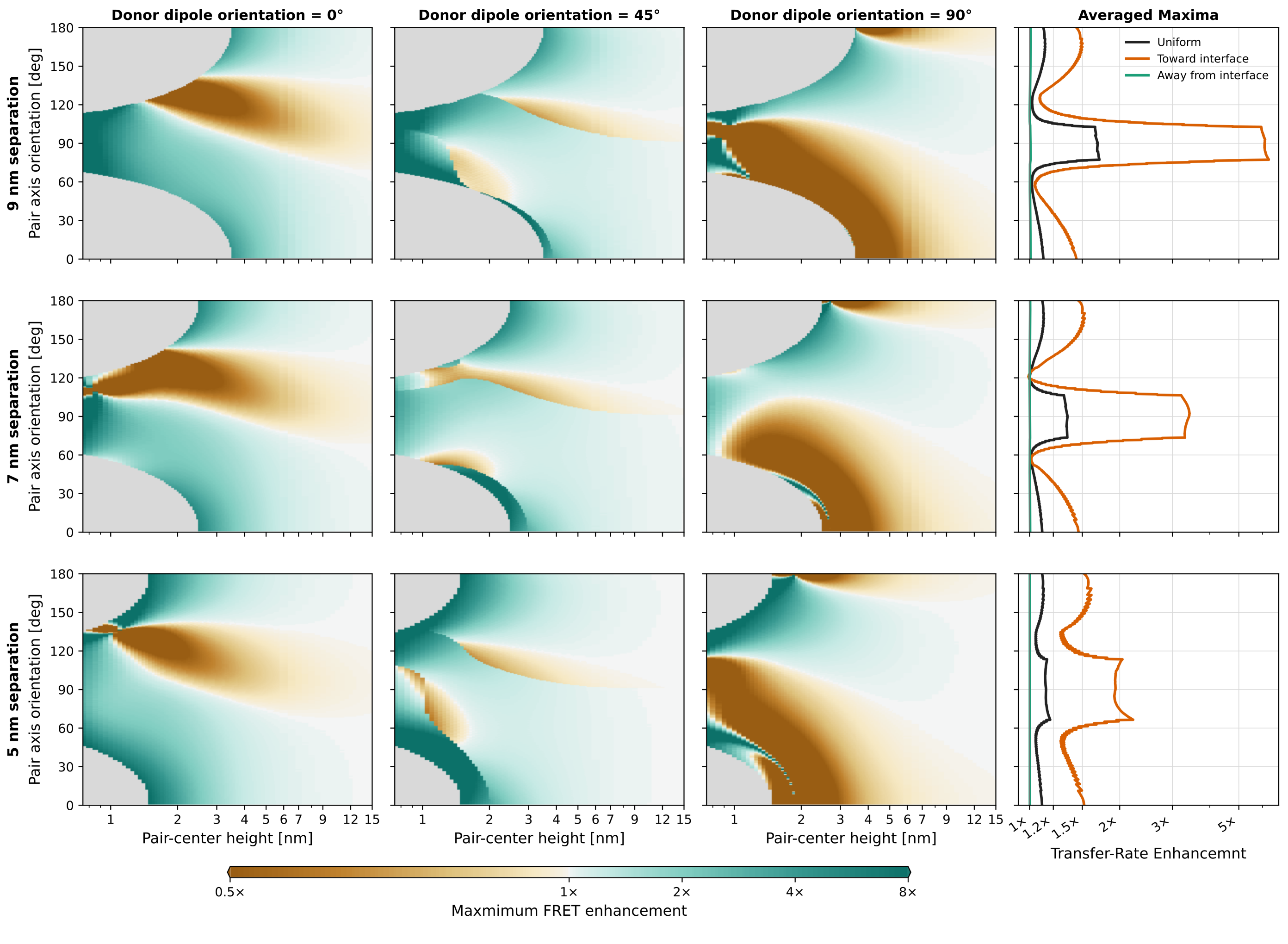} 
    \caption{Comparison of the FRET enhancement on glass and ENZ substrates for a donor-emitter pair. Each of the three rows utilizes a different value for dipole separation from 5 - 9~nm. The three left columns plot FRET enhancement as a function of the pair-axis orientation with respect to the substrate interface and the height above the substrate interface into the PVA film (log spaced). Each of the three columns represents a different value of the donor dipole orientation with respect to the substrate interface. All plots show the maximum of the FRET enhancement with respect to all remaining free degrees of orientational freedom. Note that this represents the monochromatic enhancement at 532~nm, not the full integrated overlap integral. Substantial enhancement factors are observed for systems in close proximity to the substrate and favorable configurations, and all enhancement factors approach unity beyond 15~nm height. The right column shows the height-averaged FRET enhancement profiles for each considered separation, using three spatial averaging techniques: a uniform distribution of emitters; an exponential distribution weighted toward the substrate interface with $\tau=5$nm; and a similar distribution with the same extent weighted toward the top of the PVA film away from the substrate. Note that these plots are of the height average over the maximized orientation and position degrees of freedom, which represents an upper bound rather than a true orientational average.
    }
    \label{fig:S9}
\end{figure}

\begin{figure}[h!]
    \centering
    \includegraphics[width=\textwidth]{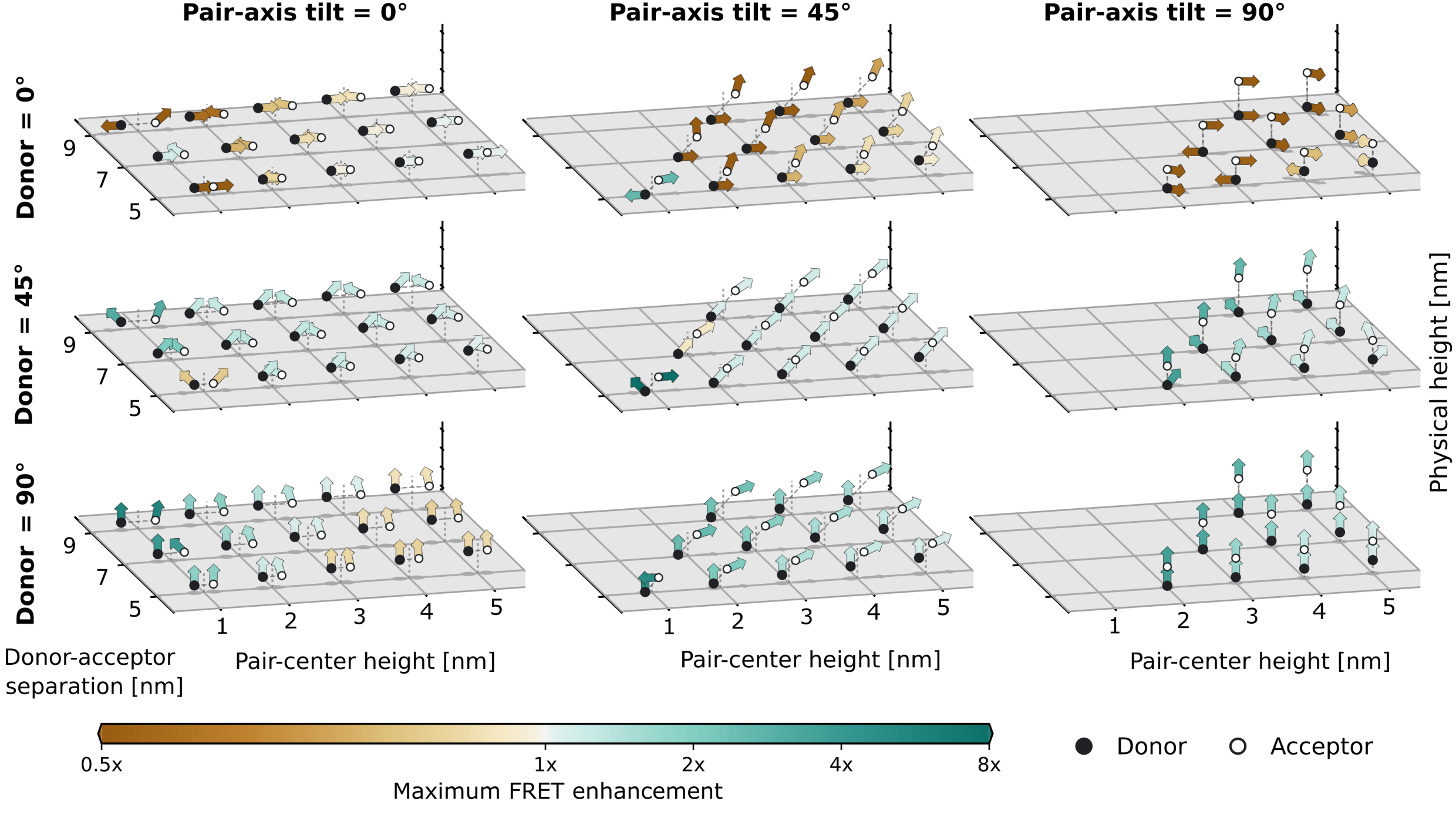} 
    \caption{Optimal dipole configurations for the donor-acceptor pair producing maximal FRET transfer enhancement for several cases of fixed pair-axis tilt (0 - 90$^\circ$) and donor dipole polar orientation (0 - 90$^\circ$), as a function of donor-acceptor pair separation and pair midpoint height above the substrate interface. Dipole positions are represented by circles, and the orientation is represented by arrows with a projection onto the substrate plane. The maximum FRET transfer-rate enhancement is shown by the arrow color.
    }
    \label{fig:S10}
\end{figure}

\begin{figure}[h!]
    \centering
    \includegraphics[width=\textwidth]{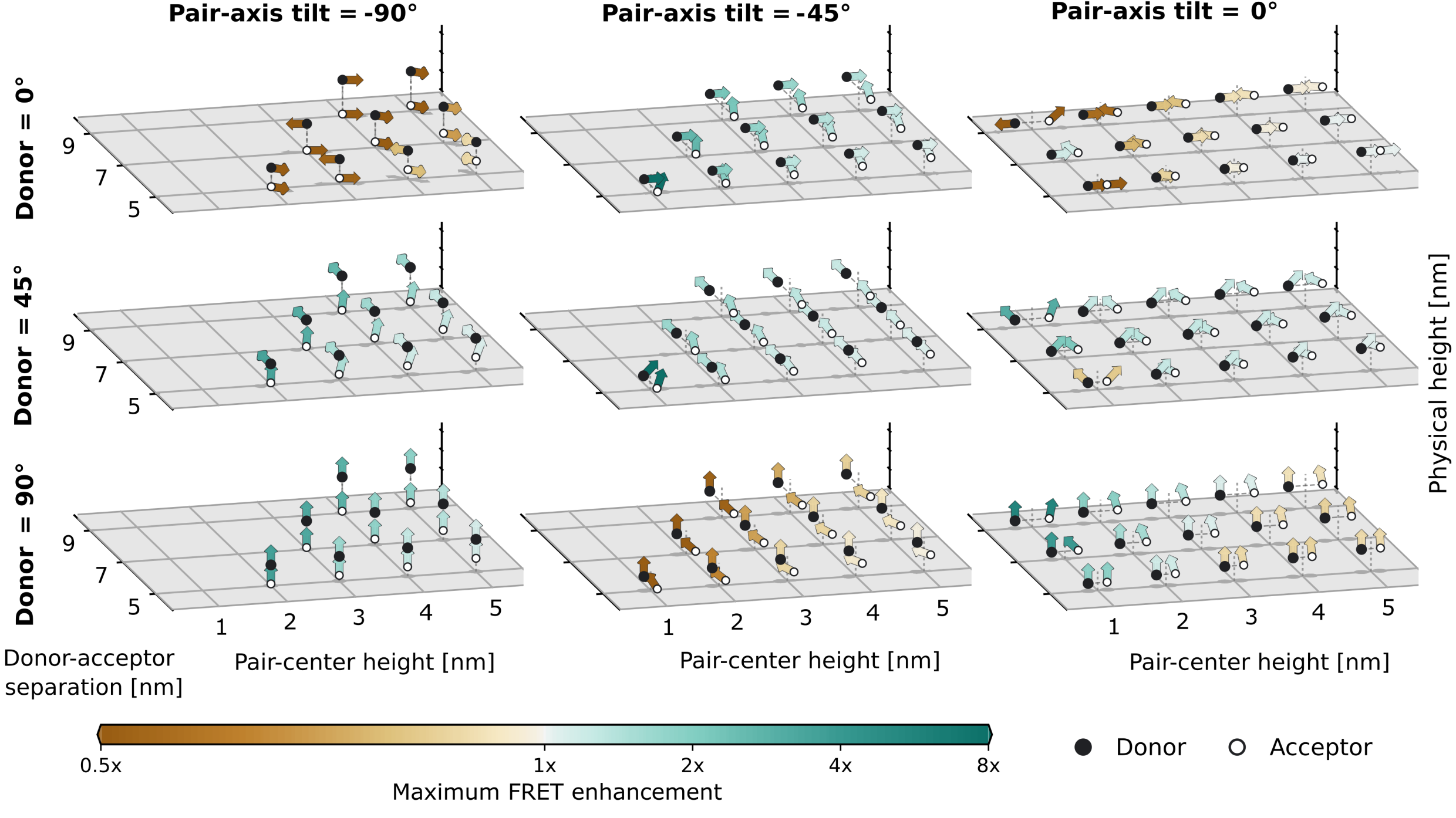} 
    \caption{Optimal dipole configurations for the donor-acceptor pair producing maximal FRET transfer-rate enhancement for several cases of fixed pair-axis tilt (-90 - 0$^\circ$) and donor dipole polar orientation (0 - 90$^\circ$), as a function of donor-acceptor pair separation and pair midpoint height above the substrate interface. Dipole positions are represented by circles, and the orientation is represented by arrows with a projection onto the substrate plane. The maximum FRET transfer-rate enhancement is shown by the arrow color.
    }
    \label{fig:S11}
\end{figure}

\subsection*{S8.6 Comparison with metallic substrates}

\begin{figure}[h!]
    \centering
    \includegraphics[width=\textwidth]{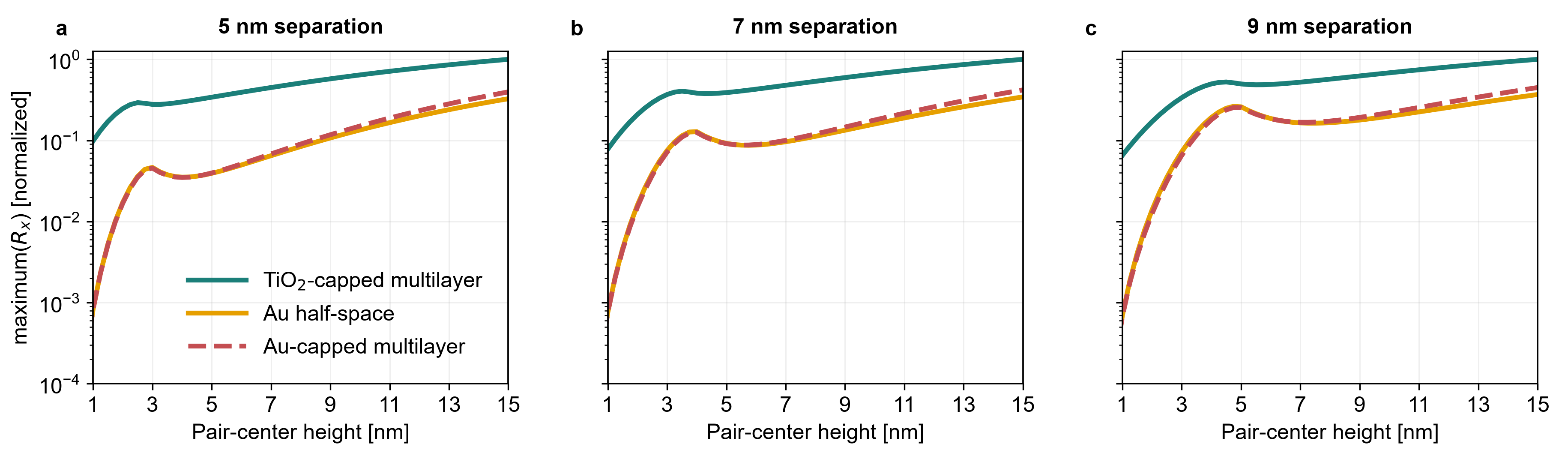} 
    \caption{Configuration-maximum FRET transfer-to-decay enhancement ratio for the TiO$_2$-capped multilayer, Au-capped multilayer, and optically thick Au reference, at donor---acceptor separations of \textbf{a}~5, \textbf{b}~7, and \textbf{c}~9~nm. At each pair-center height, $F_T/F_D$ is maximized over both dipole orientations and the allowed pair-axis orientation. All curves in each panel are normalized by the peak value of the TiO$_2$-capped multilayer.} 
    \label{fig:S12}
\end{figure}

The strongest transfer-rate enhancements occur for donor---acceptor pairs within the first few nanometers of the substrate interface. At these distances, the substrate surface termination and emitter separation from the nearest metallic layer can substantially influence both donor---acceptor coupling and competing donor decay. To examine this, we compare the TiO$_2$-terminated ENZ multilayer with an Au-terminated ENZ multilayer (reversing the layer ordering while retaining the thicknesses per material), and an optically thick Au reference. In the TiO$_2$-terminated structure, the top dielectric layer reduces direct metal-induced decay relative to the Au interfaces. 
Since emitters on plasmonic metals can exhibit strong competing donor decay pathways, such as metal quenching, transfer-rate enhancement alone is not the best metric for comparing FRET process enhancement.  Since the total decay rate can be decomposed into components, assuming a constant non-radiative rate we can define
\begin{equation}
R_X=\frac{F_{T,X}}{F_{D,X}}, \qquad
F_{T,X}=\frac{k_{T,X}}{k_{T,0}}, \qquad
F_{D,X}=\frac{k_{D,X}}{k_{D,0}}=1-\Phi_0+\Phi_0 F_{\mathrm{LDOS},X}.
\end{equation}
where $k_{T,0}$ is the FRET transfer rate and $k_{D,0}$ is the donor-only decay rate in homogeneous PVA, and $\Phi_0$ is its intrinsic fluorescence quantum yield. $R_X$ describes enhancement of the FRET transfer process odds, rather than the pure process efficiency; however, at fixed separation, maximizing $R_X$ also maximizes the FRET efficiency.

We consider the monochromatic Green tensor at 532~nm and $\Phi_0=80\%$, with $R_X$ maximized over the allowed degrees of freedom, as in the preceding analysis. We consider separations of 5, 7, and 9~nm and pair-center heights of 1--15~nm. PVA is treated as an optical half-space. As shown in \textbf{Figure~\ref{fig:S12}}, the TiO$_2$-capped multilayer gives the largest maximum ratio throughout the investigated range. The improvement over the Au-capped structures is greatest near the interface, approaching two orders of magnitude at a pair-center height of 1~nm. The similar responses of the Au-capped ENZ multilayer and thick Au reference are consistent with stronger competition from metal-induced donor decay. Note that the denominator includes both radiative and non-radiative decay components, so this ratio is not a strict measurement of metallic absorption or quenching. Also note that these  calculations omit spectral averaging, acceptor quenching, photon collection, and corrections to the metallic permittivity in the extreme near-field.

\subsection*{S8.7 ENZ multilayer discretization}

\begin{figure}[h!]
    \centering
    \includegraphics[width=0.95\textwidth]{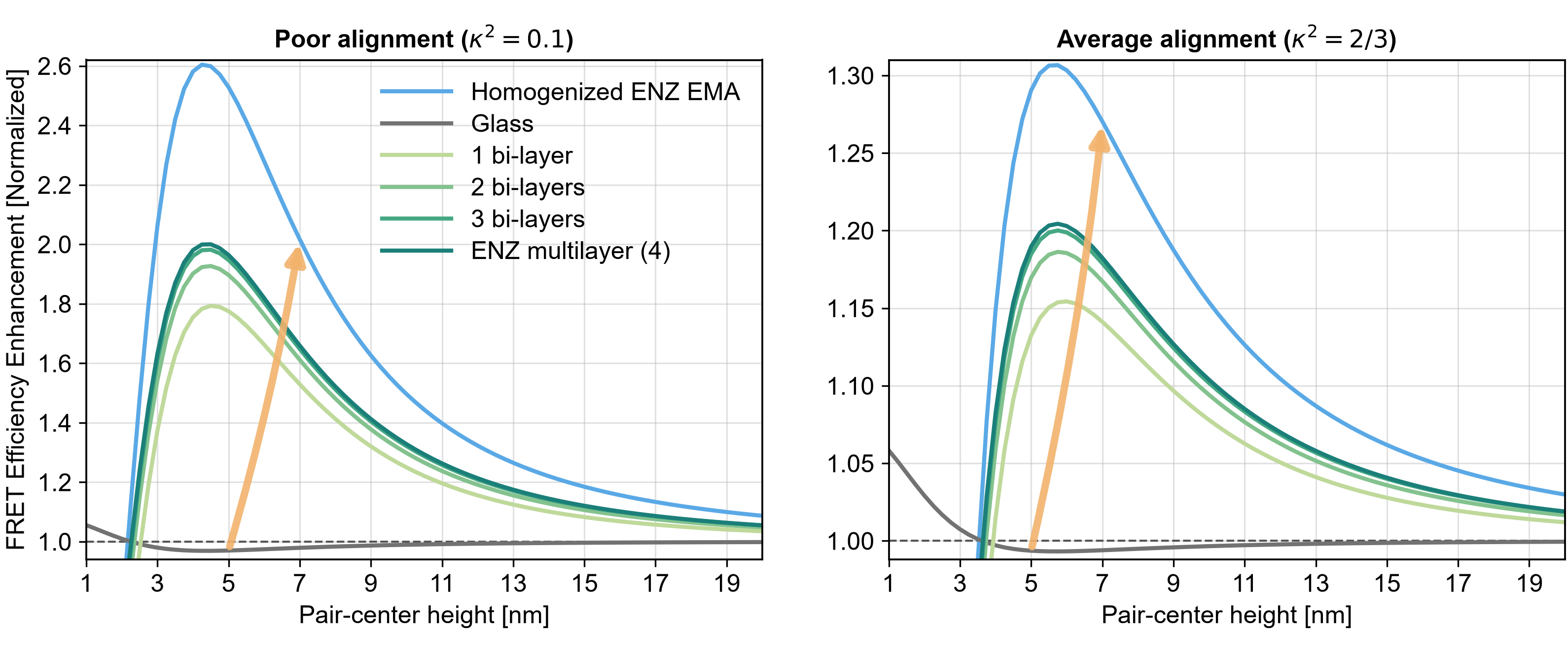} 
    \caption{Normalized FRET efficiency enhancement for a a pair of dipoles in poor ($\kappa^2=0.1$, \textbf{left}) and average ($\kappa^2=2/3$, \textbf{right}) baseline FRET alignments. Each dipole pair considers parallel dipole orientation, with the separation vector parallel to the substrate, over pair-center heights from 1--20~nm for the following substrates: a homogenized ENZ EMA (also considered in \textbf{Fig.~\ref{fig:S5}}); fused silica glass; and multilayer ENZs consisting of 1 -- 4 stacked bilayers (the manuscript condition). The orange arrow shows how the enhancement changes as the system approaches the effective ENZ.}
    \label{fig:S13}
\end{figure}

The ENZ substrate investigated in the manuscript consists of a discrete multilayer stack of metallic and dielectric layers that collectively exhibit an effective ENZ response, as observed by spectroscopic ellipsometry. However, because emitter enhancement is predicted to be maximized in the extreme near-field of the substrate -- PVA interface ($<5$~nm), finite layer discretization may become non-negligible. In this regime, the local electromagnetic environment experienced by the emitter can differ from that predicted by the homogeneous ENZ EMA. It is therefore necessary to distinguish contributions arising from the effective ENZ behavior of the metamaterial from those attributable to the specific near-field interfacial and layering effects.

In general, an emitter located in the extreme near field of a deeply subwavelength multilayer stack will feel effects from the totality of its electromagnetically accessible environment. In particular, this includes contributions from resonant phenomena associated with individual buried interfaces as well as coupled resonances mediated by multiple interfaces.
The distinction between first-interface near-field effects and collective multilayer resonance is well-established in hyperbolic metal-dielectric multilayer stacks, similar to those investigated in the manuscript. As demonstrated by Sreekanth \textit{et al.},\supercite{sreekanth2013experimental} emitters in the extreme near-field of an isolated metallic interface can access localized surface plasmon polariton (SPPs) modes, while emitters near subwavelength multilayer stacks can access multilayer bulk plasmon polariton (BPPs) modes that penetrate and propagate through the interior volume of the metamaterial and give rise to the effective hyperbolic response. In the extreme near field, high-momentum evanescent components emitted by the dipole pair couple not merely to the top dielectric-metal interface but also into volume modes supported across the buried subsurface layers. As the number of bilayers increases, the hybridization internal modes densifies, and the overall electromagnetic environment cna approach the effective ENZ response.

This point is elaborated for the case of our FRET system by considering variations on the multilayer ENZ described in the manuscript, shown in \textbf{Fig.~\ref{fig:S13}}. Two cases are considered: a pair of dipoles in poor ($\kappa^2=0.1$) and average ($\kappa^2=2/3$) baseline FRET alignments. Each dipole pair holds the dipole orientation vectors parallel, with the separation vector parallel to the substrate (both emitters at the same physical height), and considers a range of pair-center heights from 1--20~nm for the following substrates: a homogenized ENZ EMA (also considered in \textbf{Fig.~\ref{fig:S5}}); fused silica glass; and multilayer ENZs consisting of 1 -- 4 stacked bilayers (the manuscript condition). From these results, a single bilayer has a significant effect upon the FRET transfer efficiency enhancement, which describes the effect of the interface upon the FRET process since the effective ENZ properties are not present in a single bilayer. When increasing the number of bilayers, the FRET efficiency enhancement increases toward the ENZ EMA result, demonstrating the significant contribution from internal layers approaching the ENZ EMA enhancement (if the EMA is considered the "idealized ENZ behavior"). Furthermore, when considering more layers, the enhancement converges below the EMA result, demonstrating the differences present in a real multilayer system. These results show that although the microscopic near field certainly influences the FRET behavior of emitter pairs, additional bulk modes associated with the effective ENZ behavior in the far-field can have a substantial effect upon the FRET behavior in the near field. The system is fundamentally affected by both the microscopic near-field structure and the volume modes approaching the effective permittivity (ENZ behavior).

\subsection*{S8.8 Comparison with Experimental Enhancement Factors}
In the prior analysis, we have considered the change in transfer efficiency due to the presence of a substrate with respect to a fixed value in the homogeneous PVA superstrate, which avoids the poor orientational baseline conditions leading to large enhancement factors. Nevertheless, a comparison with the experimentally obtained enhancement factors requires the direct comparison between glass and ENZ substrates in the case of both the donor-only and donor-acceptor FRET pairs.  In the following analysis, the multilayer ENZ structure is utilized directly.

In \textbf{Figure~\ref{fig:S7}}, the theoretical decay rate enhancement near the interface for a single dipole (the donor-only case) is shown as a function of dipole orientation and dipole height above the substrate for three different values of the donor quantum yield ($\Phi_0$= {50, 70,90}). Although very large enhancement factors can be obtained near to the surface, the magnitude of the enhancement quickly decreases toward unity into the bulk over $\sim$30~nm. Three height-dependent averaging scenarios are considered over a 35~nm thick film region: a uniform distribution, an exponential density profile biased toward the substrate interface with $\tau=5$~nm, and an exponential density profile biased away from the surface with the same spatial dependence. The exponential profiles are meant to represent the cases of substrate-mediated attraction or repulsion. This result predicts a larger enhancement than is experimentally observed for uniformly distributed emitters near to the interface; however, this analysis only considers the total radiative and non-radiative enhancement with equal photon count weighting. In practice, the LDOS enhancement near the interface, where the greatest enhancement factors are located, can create sub-populations with significant non-radiative components associated with, for instance, emitter quenching, which do not participate equally in the experimental photon count. The radiative modal fraction proxy --- using $\mathrm{k}_{\parallel} \leq \mathrm{k}_{PVA}$ which is truly just an upper bound on the experimental photon partition --- is shown as a function of distance from the interface in \textbf{Figure~\ref{fig:S8}} for glass and ENZ substrates. This can appear as a bias in the experimental collection toward emitter populations at greater radiative fractions, which have qualitatively similar sensitivity to the apparent emitter height profile biased away from the surface. Considering the averaged enhancement profile biased away from the interface, enhancement factors of 2 - 4x are observed, in qualitative agreement with the experimental results. This is not direct evidence of substrate-mediated repulsion within the PVA matrix but is one potential explanation for the observed experimental results.

Similarly, the direct enhancement of the FRET process can be directly calculated over the glass and ENZ substrates. The FRET enhancement is highly dependent on the various orientation and positional parameters as discussed above. The results of this analysis are shown in \textbf{Figure~\ref{fig:S9}} for various separation distances and dipole orientations as a function of pair-center height above the interface, similar to the results shown in \textbf{Figure~\ref{fig:fig5}d}. Note that gray regions represent disallowed configurations due to the relative orientation. Each plot shows the maximum enhancement over all other parameters, and the colorbar is truncated due to the presence of extremely large enhancement factors, as discussed above. As with the decay rate above, the enhancement quickly decays toward unity into the bulk in the first 10~nm to 15~nm. Near the interface, there are significant regions of both enhanced and diminished enhancement, covering the experimentally obtained range. Averages of the pair center height over the maximum for all other orientational and positional parameters (except pair-axis orientation) are shown in the rightmost column of \textbf{Figure~\ref{fig:S9}} for each separation distance. The same three height-averaging conditions are considered as introduced above, namely, a uniform distribution and biasing toward and away from the interface. Note that this is not a true average over all parameters but represents an upper bound with the most favorable conditions. Considering the largest 9~nm separation, which is similar to the open-beacon condition, the height condition biased toward the interface contains transfer-rate enhancement factors consistent with all experimentally observed decay rate values --- in particular, the $\tau_1$ enhancement most associated with FRET activity.
Although this analysis supports that some emitter populations could experience ENZ-mediated FRET enhancement consistent with the experimental results, we acknowledge that it relies on assumptions about the positional and orientational parameters in the experimental system that could be verified in future work, for instance, by considering single-molecule rather than averaged ensemble experimental measurements.

\subsection{S8.9 Optimal dipole configurations}
The appearance of heterogeneous regions of substrate-mediated FRET transfer rate enhancement and distinct enhanced and quenched regions based on dipole spatial and angular orientation raises the question of which dipole configurations obtain maximal enhancement. Furthermore, under constrained orientation conditions, there exists a set of optimized possible configurations, which is determined by the specific field engendered by the presence of the substrate boundary conditions. In the case of the ENZ substrate, these optimal dipole configurations are given in \textbf{Figures~\ref{fig:S10}} and \textbf{\ref{fig:S11}} as a function of donor-acceptor pair separation and pair midpoint height above the substrate interface for several cases of fixed pair-axis tilt and donor dipole polar orientation. These results are an extension of the main text dipole configurations, which were maximized over these potential configuration spaces.

For large distances from the substrate, the optimal condition is collinear dipoles oriented normal to the substrate plane. However, substantial enhancement can be obtained for non-collinear configurations in other regimes and substrate separation distances. The presence of non-collinear maxima is partially driven by configurations with disallowed dipole positions, which would physically place an emitter below the substrate; however, in constrained sub-optimal configurations, for instance, at a pair-axis angle of 45$^\circ$, the substrate-mediated FRET enhancement can still be substantial. These results can provide a blueprint for future substrate-mediated FRET enhancement systems, where implementing controls on the orientational configurations of emitter pairs can significantly change the magnitude of the enhancement.

% Print the SI bibliography (starts numbering from [1])
\printbibliography[heading=subbibliography, title={Supporting References}]

\end{refsection}

\end{document}